\documentclass[twocolumn,
amsmath,amssymb,amsfonts, a4paper,aps,
citeautoscript,superscriptaddress,footinbib,prb]{revtex4-2}
\usepackage{times}
\usepackage{graphicx}
\usepackage{float}
\usepackage{latexsym,amsmath,amssymb,bm,euscript}
\usepackage{color}
\usepackage{epstopdf}
\usepackage[colorlinks=true,linkcolor=blue,citecolor=blue,urlcolor=blue]{hyperref}
\usepackage{hyperref}
\usepackage{subfig}
\usepackage{multirow}
\usepackage{caption}
\usepackage{lipsum}
\newcommand{\kagome}{Kagom$\acute{\text{e}}$ }

\begin{document}

\title{Ground State of $\mathrm{SU}\left(3\right)$ spin model on the checkerboard lattice }
\author{Jun-Hao Zhang}
\email{jhzhang16@fudan.edu.cn} %
\affiliation{Department of Physics and State Key Laboratory of Surface Physics, Fudan University, Shanghai 200433, China}
\author{Jie Hou}%
\affiliation{Department of Physics and State Key Laboratory of Surface Physics, Fudan University, Shanghai 200433, China}
\author{Jie Lou}%
\email{loujie@fudan.edu.cn}
\affiliation{Department of Physics and State Key Laboratory of Surface Physics, Fudan University, Shanghai 200433, China}
\author{Yan Chen}
\email{yanchen99@fudan.edu.cn}
\affiliation{Department of Physics and State Key Laboratory of Surface Physics, Fudan University, Shanghai 200433, China}
\affiliation{Shanghai Branch, Hefei National Laboratory, Shanghai 201315, China}
%\thanks{ }

\date{\today}% It is always \today, today,
\begin{abstract}
Geometric frustration in quantum spin systems can lead to exotic ground states.
In this study, we investigate the $\mathrm{SU}(3)$ spin model on the checkerboard lattice to explore the effects of frustration arising from its point-connected $(N+1)$-site local structure.
We employ density matrix renormalization group (DMRG) and exact diagonalization (ED) techniques to determine the ground state properties.
Our results reveal the absence of both 3-sublattice antiferromagnetic order and valence cluster solid order.
Instead, we identify ground states with bond stripe patterns sensitive to boundary conditions and system size, comprising staggered singlet arrays and uniform flat stripes.
Notably, these stripes are relatively decoupled, and similar patterns can be reconstructed in quasi-one-dimensional ladders.
These findings suggest that geometric frustration drives the system toward a mixed phase, combining characteristics of spin-liquid and valence cluster solid states, providing new insights into the behavior of frustrated quantum spin systems.

\end{abstract}

\maketitle

%%%%%%%%%%%
\section{Introduction}

Over the past few decades, frustrated magnets have garnered significant interest from researchers. In these systems, the suppression of conventional magnetic order allows quantum fluctuations to dominate the physics. Consequently, complex quantum phenomena—including quantum criticality, topological orders, and long-range entanglement—emerge in frustrated magnets. Exotic quantum phases, such as the quantum spin liquid, have been discovered in frustrated spin systems.
In theoretical models, frustrated magnets exhibit a rich variety of properties that depend on both the source of frustration and the intrinsic characteristics of the system.

Systems exhibiting $\mathrm{SU}(N)$ symmetry are of particular interest.
Alkaline-earth-like atoms trapped in an optical lattice, such as $^{179}\rm{Yb}$ and $^{87}\rm{Sr}$, can be used to realize this symmetry~\cite{stellmer_detection_2011,stellmer_bose-einstein_2009,de_escobar_bose-einstein_2009,kitagawa_two-color_2008}.
Due to the presence of two outermost electrons, these atoms exhibit no electronic magnetic moment.
Instead, the interactions of the hyperfine structure can be finely tuned by orbital Feshbach resonance~\cite{hofer_observation_2015,pagano_strongly_2015,zhang_orbital_2015}, enabling the realization of $\mathrm{SU}(N)$ symmetry in these systems~\cite{cazalilla_ultracold_2009,gorshkov_two-orbital_2010}.
Various values of $N$ and optical lattice geometries have been experimentally realized and explored~\cite{taie_su6_2012,hofrichter_direct_2016,ozawa_antiferromagnetic_2018,tusi_flavour-selective_2022,taie_observation_2022}.
Recent advancements have enabled quantum simulations of ultracold two-dimensional Hubbard and Heisenberg-type models~\cite{muller_state_2021,yamamoto_engineering_2024}.

The $\mathrm{SU}(N)$ Fermi Hubbard model is a natural generalization of the conventional Fermi Hubbard model, characterized by a hopping parameter $t$ and an on-site repulsion $U$~\cite{honerkamp_ultracold_2004}.
This model has been intensively studied over the past decades~\cite{sotnikov_magnetic_2014,sotnikov_critical_2015,hafez-torbati_artificial_2018,chen_synthetic-gauge-field_2016}.
The case $N=3$ is of particular interest due to its connection to the gauge symmetry in quantum chromodynamics and its computational tractability.
Recent studies on the square lattice reveal a metal-insulator transition and quantum antiferromagnetic (AFM) ordering with different periodicities near integer fillings~\cite{feng_metal-insulator_2023,ibarra-garcia-padilla_metal-insulator_2023}.
In one-dimensional systems, trionic molecular superfluidity and topological phases have been observed~\cite{assaraf_metal-insulator_1999,capponi_phases_2016}.
In contrast, the strongly correlated limit of the $\mathrm{SU}(N)$ Fermi Hubbard model leads to the $\mathrm{SU}(N)$ $t$-$J$ model, which exhibits unique properties~\cite{he_six-component_2023,schlomer_subdimensional_2024,zhang_ground_2024}.

The spin model can be regarded as an effective approximation of the $\mathrm{SU}(N)$ Fermi Hubbard model at $1/N$-filling ($\langle n \rangle = 1$) in the large-$U$ limit~\cite{hermele_topological_2011}.
In this regime, Mott physics suppresses charge fluctuations, and the effective interactions between spin degrees of freedom can be derived via second-order perturbation theory.

Numerous novel physical phenomena have been observed in studies of $\mathrm{SU}(N)$ spin models in two-dimensional systems~\cite{szasz_phase_2022,kaneko_ground-state_2024}.
A comprehensive phase diagram of the $\mathrm{SU}(N)$ Heisenberg model has been derived using large-$N$ mean-field approximations~\cite{hermele_topological_2011,yao_topological_2021,yao_intertwining_2022,chen_synthetic-gauge-field_2016}.
This phase diagram features a chiral spin liquid phase at large $N$, a valence cluster solid at small $N$, and stripe states with flux at intermediate $N$.
Tensor network calculations on certain $\mathrm{SU}(N)$ spin models with chiral interactions have further confirmed the presence of the chiral spin liquid phase~\cite{chen_abelian_2021,niu_chiral_2024,xu_phase_2023}.
For the $N=3$ case on a square lattice, multiple studies predict a three-sublattice order~\cite{toth_three-sublattice_2010,bauer_three-sublattice_2012,lauchli_quadrupolar_2006,lauchli_erratum_2006},
whereas density matrix renormalization group calculations suggest the possibility of a fragile nematic spin liquid~\cite{hu_density_2020,zhang_fragility_2021,herviou_even-odd_2023}.
Exploiting the non-Abelian symmetry inherent in these models, numerous state-of-the-art numerical techniques have been developed, yielding results with remarkable scale and precision~\cite{botzung_exact_2024,weichselbaum_qspace_2024}.

In $\mathrm{SU}(N)$ spin models, the local frustration structure is fundamentally different from that in conventional $\mathrm{SU}(2)$ systems.
For $N=3$, the local frustrated motif transitions from a triangular arrangement to a tetrahedral one.
The checkerboard lattice examined in this paper can be viewed as a two-dimensional assembly of locally frustrated units that share corners, and it is equivalent to a line graph lattice. In this context, the line graph lattice represents the link-point dual of the original lattice, analogous to the \kagome lattice.
Tight-binding models defined on such line graph lattices exhibit a high degree of degeneracy in their energy spectra, which gives rise to flat energy bands.
This characteristic, in turn, leads to various intriguing phenomena, including nontrivial topological phases, band ferromagnetism, and superconductivity~\cite{sun_nearly_2011,yoshioka_frustration_2008,fujimoto_geometrical-frustration-induced_2002,katsura_nagaoka_2013,tamura_ferromagnetism_2019,tamura_ferromagnetism_2021}.
Experimentally, this lattice has been realized in monolayer $\mathrm{Cu_{2}N}$~\cite{hu_realization_2023}.
Moreover, the conventional spin model on the checkerboard lattice has been the subject of extensive theoretical investigation~\cite{canals_square_2002,fouet_planar_2003,bernier_planar_2004,chen_exact_2006,starykh_anisotropic_2005,bishop_frustrated_2012,li_ground-state_2015,capponi_numerical_2017,wildeboer_exact_2020}, and this lattice has also been explored for the $\mathrm{SU}(4)$ spin model as a suitable platform for realizing a four-site simplex valence cluster solid (VCS) state~\cite{corboz_simplex_2012}.

In this paper, we investigate the ground state of the frustrated $\mathrm{SU}(3)$ spin model on a checkerboard lattice.
Using exact diagonalization (ED) and density matrix renormalization group (DMRG) calculations, we determine the ground state of a Hamiltonian that continuously connects the square lattice to the checkerboard lattice via next-nearest-neighbor (NNN) interactions.
We observe a quantum phase transition as the NNN interactions evolve continuously.
For the ED results, we find a change in the ground state degeneracy, while the DMRG results show a sudden reduction in the correlation length.
Furthermore, we compare various low-energy states computed using DMRG and conclude that a bond stripe pattern is favored.
Next, we investigate the nature of the stripe states and find that the bond stripes are nearly decoupled.
We characterize the single-stripe subsystems and reproduce the stripe pattern in quasi-one-dimensional ladders.
Finally, we summarize our findings and discuss their relation to previous works and low-energy effective theories.

The organization of our paper is as follows.
In Section~\ref{sec:m_m}, we introduce the $\mathrm{SU}(3)$ Heisenberg model on the checkerboard lattice and the parameter settings for the DMRG simulations.
In Section~\ref{sec:numer_result}, we present our numerical results, including various correlations and different local quantities.
In Section~\ref{sec:stripe_pattern}, we study the nature of stripe states and reproduce the stripe in quasi-one-dimensional ladders.
The final section, Section~\ref{sec:sum}, contains the summary and discussion.

\section{Model}\label{sec:m_m}

\subsection{$\mathrm{SU}\left(3\right)$ symmetric Hamiltonian}\label{sub:su3_ham}
The two-site $\mathrm{SU}(3)$ symmetric spin Hamiltonian is constructed by setting $N=3$ in the general $\mathrm{SU}(N)$ Heisenberg exchange Hamiltonian
\begin{equation}\label{eq:local_ham}
\mathcal{H}_{ij}^{\mathrm{SU}(N)} = J P_{ij} = J S_{i}^{\alpha \beta} S_{j}^{\beta \alpha},
\end{equation}
where $P_{ij}$ exchanges the local states on the two sites, and $S_{i}^{\alpha \beta} = c^{\dagger}_{i \alpha} c_{i \beta}$ are the $\mathrm{SU}(N)$ spin operators in terms of fermion operators. The indices $\alpha$ and $\beta$ sum over the $N$ spin flavors.
In the following, we focus on the case corresponding to the fundamental irreducible representation (irrep), described by the Young diagram $\Box$.
This interaction has several conserved quantum numbers. The total spin components $L^{3,8}$ and the second Casimir $C^2$ are conserved quantities. For example, the $L^3$ component can be expressed as the Gell-Mann generator $\hat{L}^{3}_{i} = \left(\hat{n}_{i}^{\alpha=1} - \hat{n}_{i}^{\alpha=2}\right)/{\sqrt{2}}$. These conserved quantum numbers can accelerate the numerical calculations.

\subsection{Checkerboard Lattice}\label{sub:cb_lat}
\begin{figure}[htbp]
\includegraphics[width=0.20\textwidth]{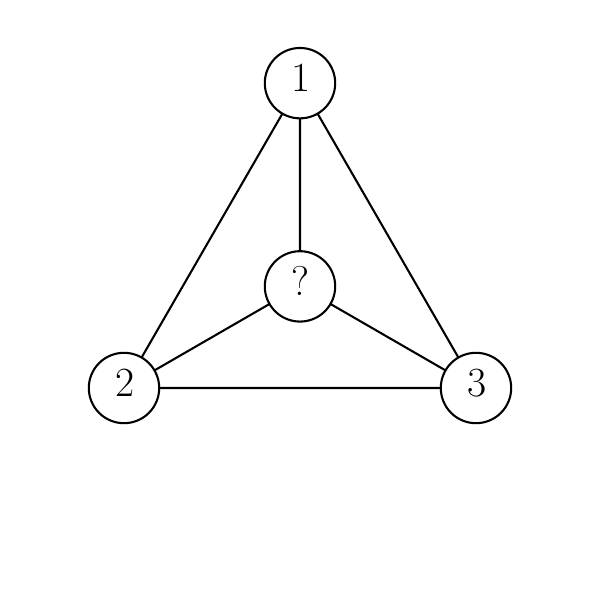}\label{fig:frustration_su3}
\includegraphics[width=0.25\textwidth]{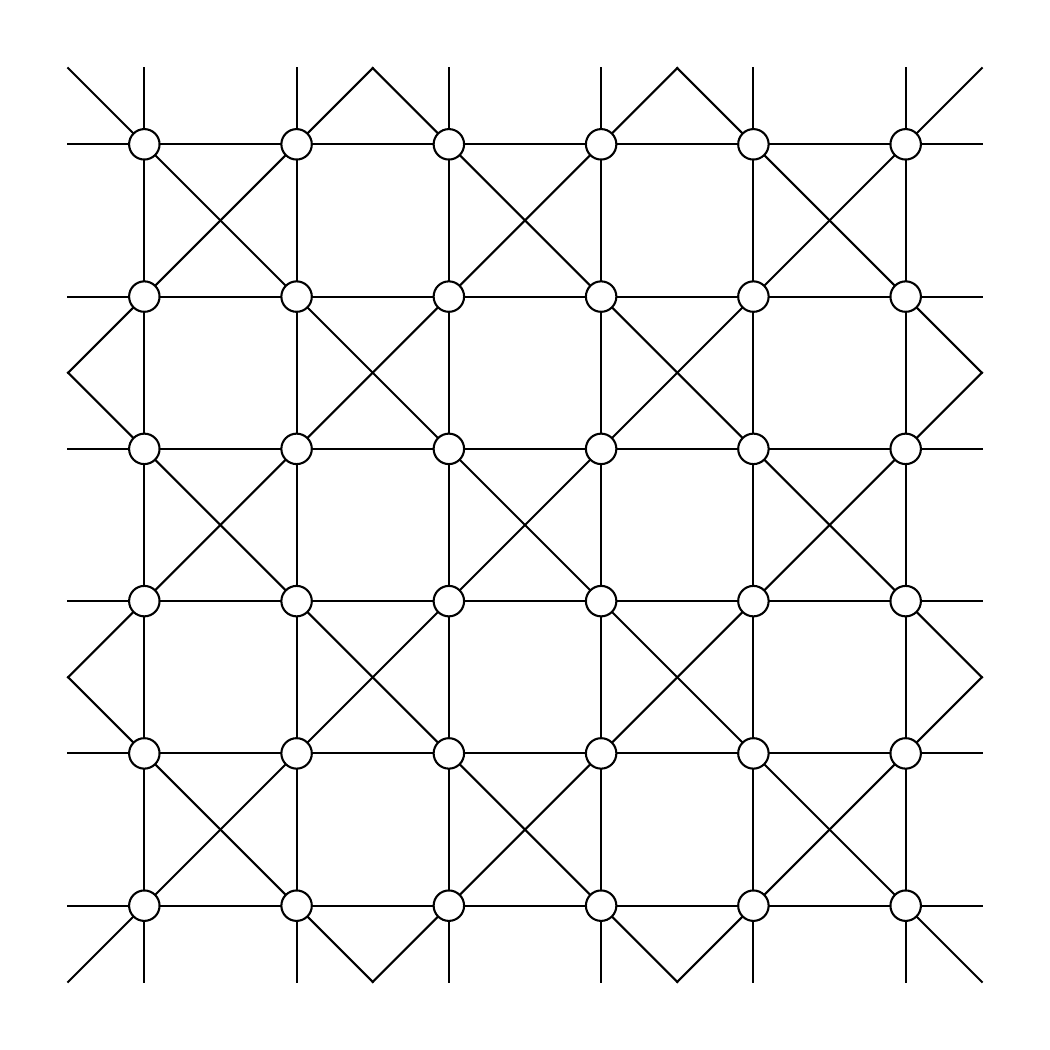}\label{fig:checkerboard_lattice}
\centering
\caption{Local frustration structure for $\mathrm{SU}\left(3\right)$ fundamental irreps (left panel) and checkerboard lattice (right panel).%
}\label{fig:frus_checker}
\end{figure}

Geometric frustration in conventional spin systems is often illustrated using triangular structures~\cite{balents_spin_2010, lacroix_introduction_2011}.
In models where a singlet requires two lattice sites to form, such as spin-$\frac{1}{2}$ systems, the \kagome lattice is considered frustrated.
This lattice consists of triangles that share only a single corner with neighboring triangles.
For $\mathrm{SU}(3)$ symmetry, singlet formation requires three fundamental irreducible representations.
Specifically, a singlet occupies three sites in the Hamiltonian given by Eq.~(\ref{eq:local_ham}).
Consequently, the notion of local frustration must be generalized to involve four connected sites (FIG.~\ref{fig:frus_checker}).
By repeating this structure such that each unit shares only one corner with its neighbors, we obtain the 3D pyrochlore lattice.
The checkerboard lattice also satisfies these structural requirements and is sometimes referred to as the planar pyrochlore lattice (FIG.~\ref{fig:frus_checker}).
These lattices can be classified as line graphs of different parent lattices with a flat upper energy band.

To relate with the previous results for square lattice, checkerboard lattice is regarded as a square lattice with next-nearest-neighbor (NNN) interactions on half of the plaquette.
The Hamiltonian can be expressed as
\begin{equation}\label{eq:checker_ham}
H_{\text{Checkerboard}}^{\mathrm{SU}\left(N\right)} = \sum_{\left<i,j\right>} J S_{i}^{\alpha \beta} S_{j}^{\beta \alpha} + \sum_{\left<\left<i,j\right>\right>}^{\text{half plaquette}} J_{\times} S_{i}^{\alpha \beta} S_{j}^{\beta \alpha}.
\end{equation}

We can interpolate between the square lattice and the checkerboard lattice by varying $J_{\times}$ from $0$ to $J$.
This approach allows us to investigate whether the ground states on the square lattice and the checkerboard lattice belong to the same phase.
For simplicity, we set $J \equiv 1$ to define the overall energy scale.
The unit cell of the checkerboard lattice is twice the size of that in the square lattice, containing two sites.
The basis vectors of the checkerboard lattice are rotated by $\pi/4$ relative to those of the square lattice, and their lengths are $\sqrt{2}a$, where $a$ is the lattice constant of the original square lattice.

\section{Numerical Results}\label{sec:numer_result}
We use ED and DMRG to study the ground state of this model.
For the ED computations, we employ the QuSpin package~\cite{weinberg_quspin_2017, weinberg_quspin_2019}.
Periodic boundary conditions are applied in both directions during the ED simulations.
The DMRG simulations are performed using the ITensor library~\cite{fishman_itensor_2022}.
A maximum bond dimension of $D_{\text{max}} = 6000$ is maintained during the DMRG simulations.
The calculations typically involve around 100 sweeps, with the largest truncation errors of the final wavefunctions on the order of $10^{-5}$.
Global conservation of the $\mathrm{SU}(3)$ spin component is preserved as a good quantum number to save computational resources.

\subsection{ED Result}\label{sub:ed_result}

\begin{figure}[htbp]
  \centering
  \includegraphics[width=0.45\textwidth]{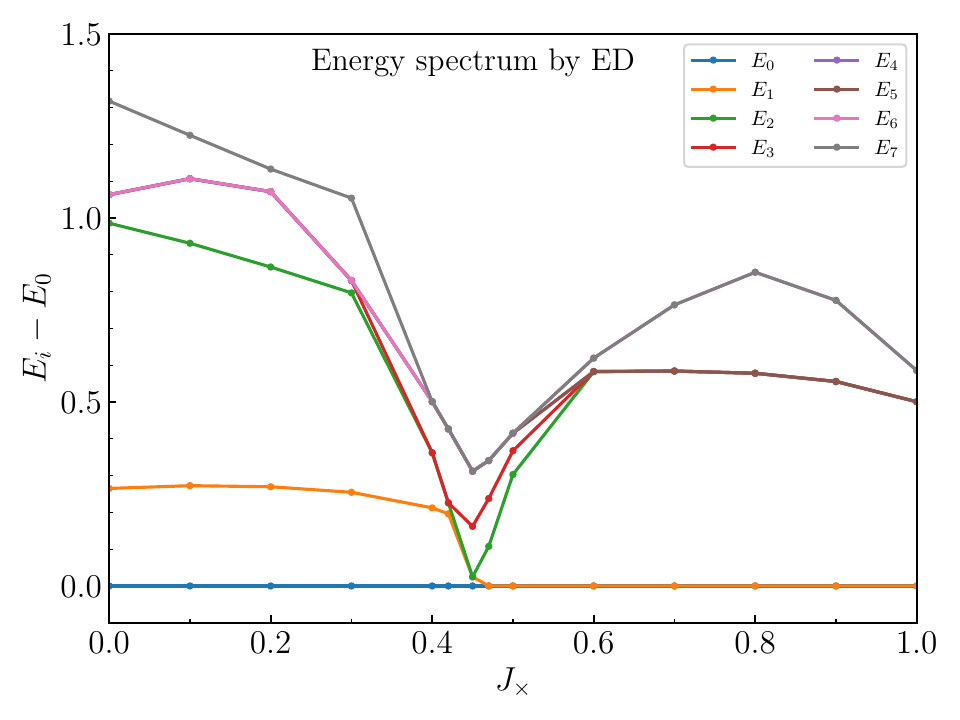}
  \caption[ed_spectrum]{%
      Low energy spectrum of $3\times 3$ system.}%
  \label{fig:ed_spec}
\end{figure}

We calculate the energy spectrum of several of the lowest-energy states for a system with $3 \times 3$ unit cells, containing 18 sites.
The next-nearest-neighbor (NNN) interaction strength, $J_{\times}$, varies from 0 to 1 to capture distinct phases.
The spectrum is shown in Fig.~\ref{fig:ed_spec}.
A phase transition occurs near $J_{\times} = 0.45$, where the energy levels intersect.
The ground state is unique for $J_{\times} < 0.45$ and becomes twofold degenerate for $J_{\times} > 0.45$.

\begin{figure}[htbp]
  \centering
  \subfloat[$J_{\times}=0.10$  \centering \label{fig:ed_pattern_a}]{
      \includegraphics[width=0.13\textwidth]{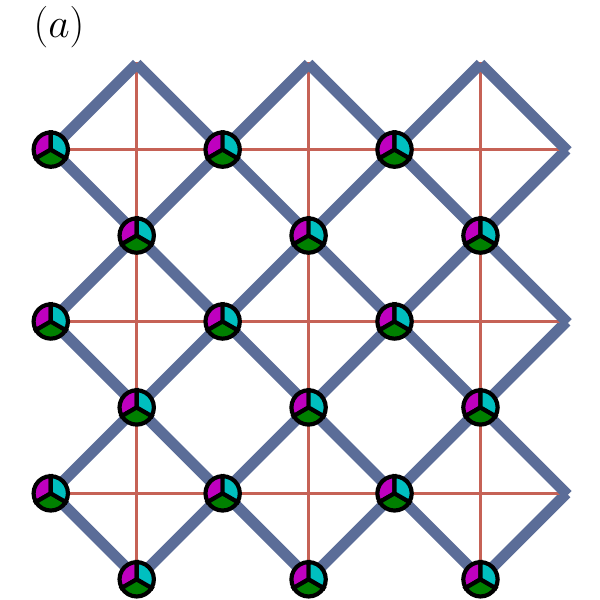}}
  \hspace{0.01\textwidth}
  \subfloat[$J_{\times}=0.90$  \centering \label{fig:ed_pattern_b}]{
      \includegraphics[width=0.13\textwidth]{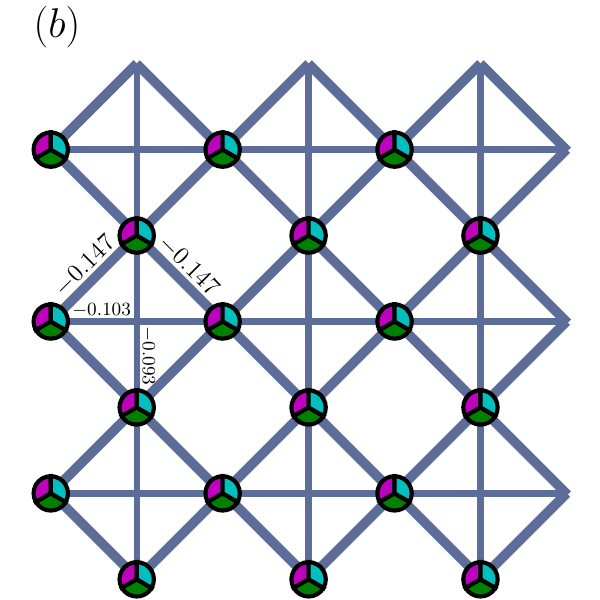}}
  \hspace{0.01\textwidth}
  \subfloat[$J_{\times}=1.18$  \centering \label{fig:ed_pattern_c}]{
      \includegraphics[width=0.13\textwidth]{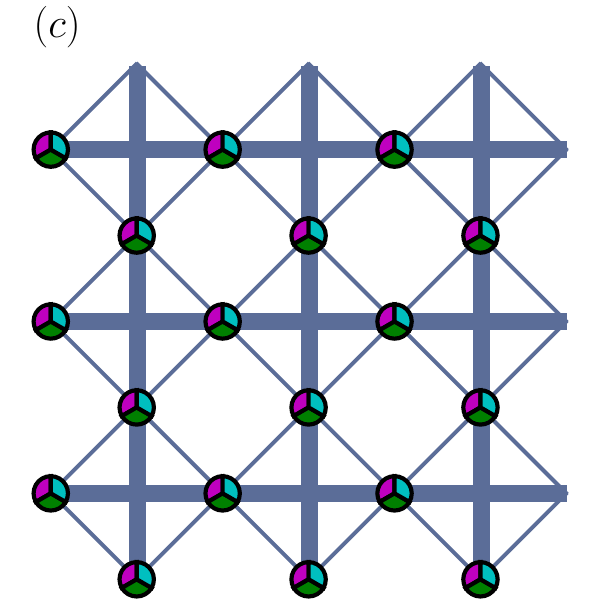}}
  \caption[ed_pattern]{%
      Real-space coupling patterns of the ground states in two phases.
      (a) unique ground state when $J_{\times}=0.10$; (b) one of two degenerate ground states when $J_{\times}=0.90$; (c) unique ground state when $J_{\times}=1.18$.
      The pie charts represent the local spin component on each site. The width of the lines represents the strength of the coupling.
      Its color takes blue and red for AFM and ferromagnetic coupling.
      }\label{fig:ed_pattern}
\end{figure}

This twofold degeneracy can be lifted by reducing the lattice symmetry from $C_4$ to $C_2$.
Specifically, applying different values of $J_{\times}$ along various directions splits the twofold-degenerate ground states.
This allows us to extract the coupling strengths along the interaction bonds.
The two degenerate ground states exhibit slight differences in the NNN couplings along different directions.
The local spin component $n_{i}^{\alpha}$ and coupling strength $B_{ij} = S_{i}^{\alpha \beta} S_{j}^{\beta \alpha}$ are shown in Fig.~\ref{fig:ed_pattern}.
The NNN coupling strengths $B_{\text{NNN}}$ along two directions differ slightly, with values of -0.104 and -0.093.
As a result, the degenerate ground states break symmetry, potentially resembling a nematic spin liquid~\cite{zhang_fragility_2021}.
Furthermore, when $J_{\times}$ exceeds 1.15, the ground state transitions into a direct product of six singlets aligned along the $x$ and $y$ directions.
This occurs because the finite system size allows singlets to form across the boundary, a phenomenon that would not be permitted in the thermodynamic limit.

\subsection{DMRG Result}\label{sub:dmrg_result}
The ED results indicate a phase transition between square and planar pyrochlore lattices.
However, ED introduces significant finite-size effects and offers limited information about correlations.
To address these limitations, we employ the DMRG algorithm to study the ground state in larger system sizes.

The DMRG calculations are performed on cylindrical geometries, employing open boundary conditions along the $x$ direction and periodic boundary conditions along the $y$ direction.
The system size is denoted as $N_x \times N_y$ unit cells, comprising a total of $N_{\text{tot}} = 2N_x N_y + N_y$ sites.
The additional $N_y$ sites correspond to boundary sites on one of the open boundaries.
We examine multiple system sizes, with $N_x$ varying from 7 to 10 and $N_y$ from 4 to 6.
The system sizes are chosen such that $N_{\text{tot}}$ is a multiple of 3.

To complement the ED results, we investigate the evolution of the ground state as $J_\times$ varies from 0 to 1.
We measure the spin correlation function
\begin{equation}\label{eq:spin_correlation}
    S\left(r\right)=\left<\hat{L}_{i}^{3}\hat{L}_{i+\vec{r}}^{3}\right>.
\end{equation}

\begin{figure}[htbp]
    \centering
    \subfloat[\label{fig:spin_corre_Jx} \centering ]{
      \includegraphics[width=0.238\textwidth]{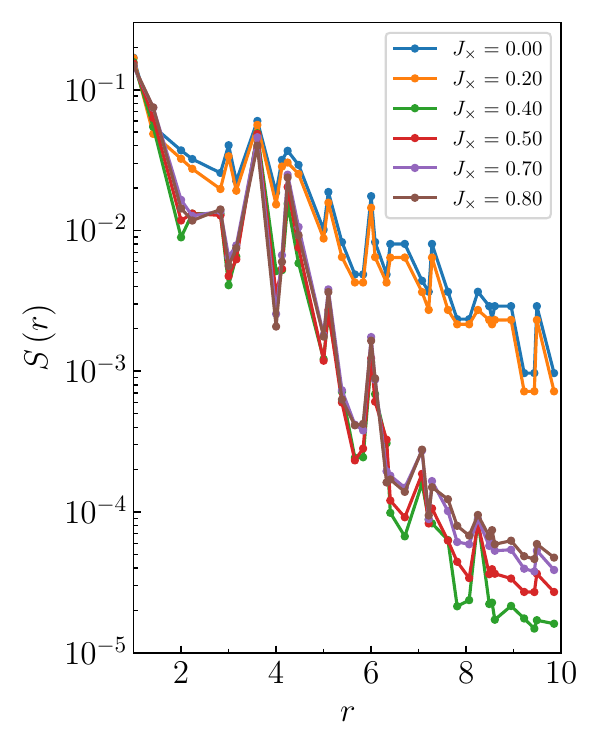}}
    \subfloat[\label{fig:spin_corre_fourier} \centering ]{
      \includegraphics[width=0.24\textwidth]{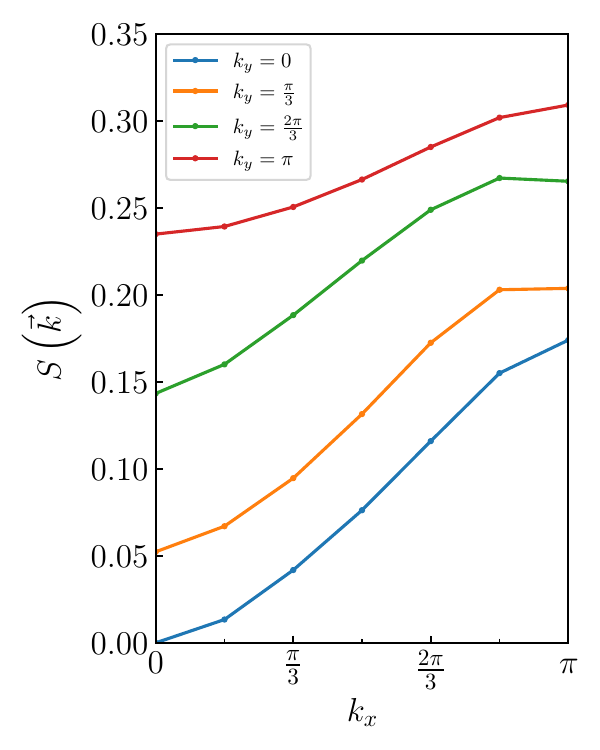}}
    \caption[spin correlation vs. Jx]{%
        Spin correlation
        $\left(a\right)$ Spin correlation with different $J_\times$, where the correlations decay faster in the strong NNN interaction case; %
        % $\left(b\right)$ Decay behaviors of spin correlation along the side of the plaquette in $30\times 4$ square size;%
        $\left(b\right)$ Fourier transform of spin correlation has no significant singularity.
        }\label{fig:exclude_3sub}
\end{figure}

Different behaviors are observed for weak and strong NNN interactions~(Fig. \ref{fig:spin_corre_Jx}).
In the strong interaction case, the correlation length is notably shorter compared to the weak interaction case.
This observation suggests a phase transition between the ground states of square and planar pyrochlore lattices.

The shorter spin correlation length suggests the breakdown of the three-sublattice AFM order.
The Fourier transform of the spin correlation function is presented in Fig.~\ref{fig:spin_corre_fourier}.
Its smoothness and underlying quadratic behavior near the origin imply the presence of a finite spin gap and the absence of long-range spin order.
We also examine planar pyrochlore lattice systems configured in a square lattice geometry (details in Supplemental Materials).
The spin correlation function along the plaquette's edge exhibits a short correlation length of $r_S = 0.91a$.
Additionally, we attempt to induce three-sublattice AFM order and valence bond (or cluster) solid order by applying corresponding fields at the boundaries.
However, our results do not support the existence of these orders (details in Supplemental Materials).

%%%
\subsection{Real-Space Valence Bond Coupling}\label{sub:valence_bond}
\begin{figure}[htbp]
  \centering
  \subfloat[$\left(7,4\right)$ Stripe Structure \centering \label{fig:local_unitcell_4x7_a}]{
      \includegraphics[width=0.45\textwidth]{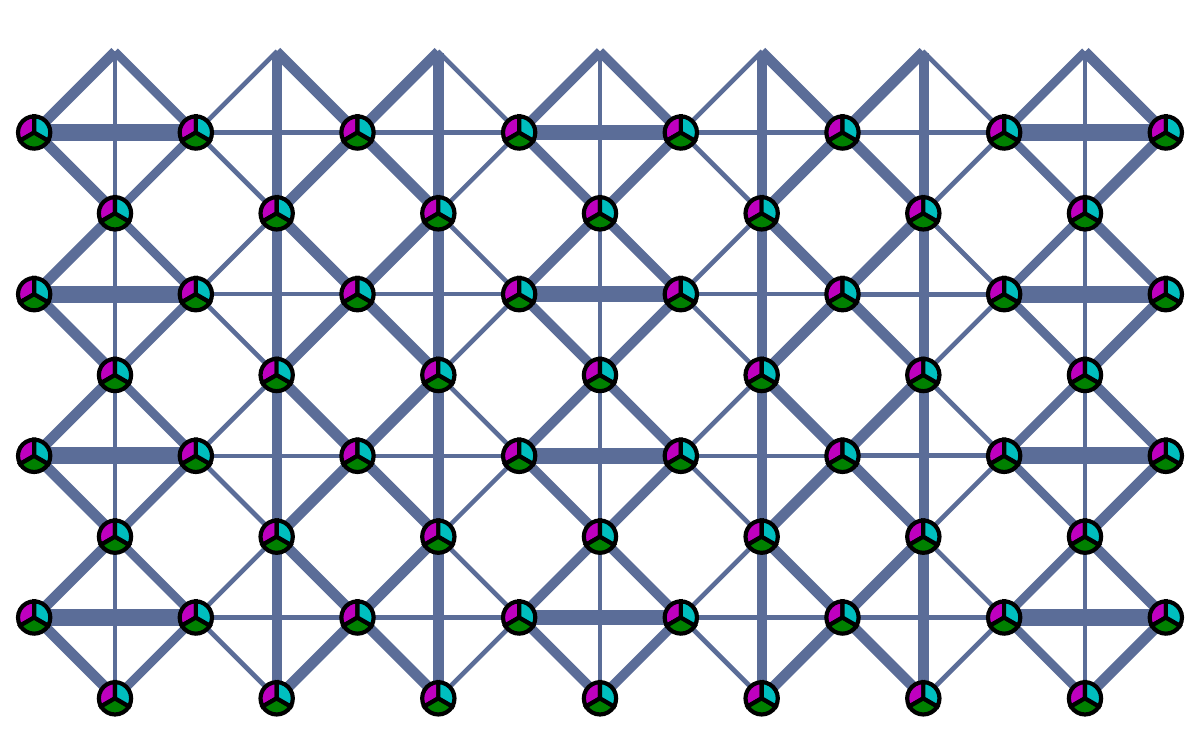}}
  \hspace{0.01\textwidth}
  \subfloat[$\left(7,4\right)$ Local Minimum \centering \label{fig:local_unitcell_4x7_b}]{
      \includegraphics[width=0.45\textwidth]{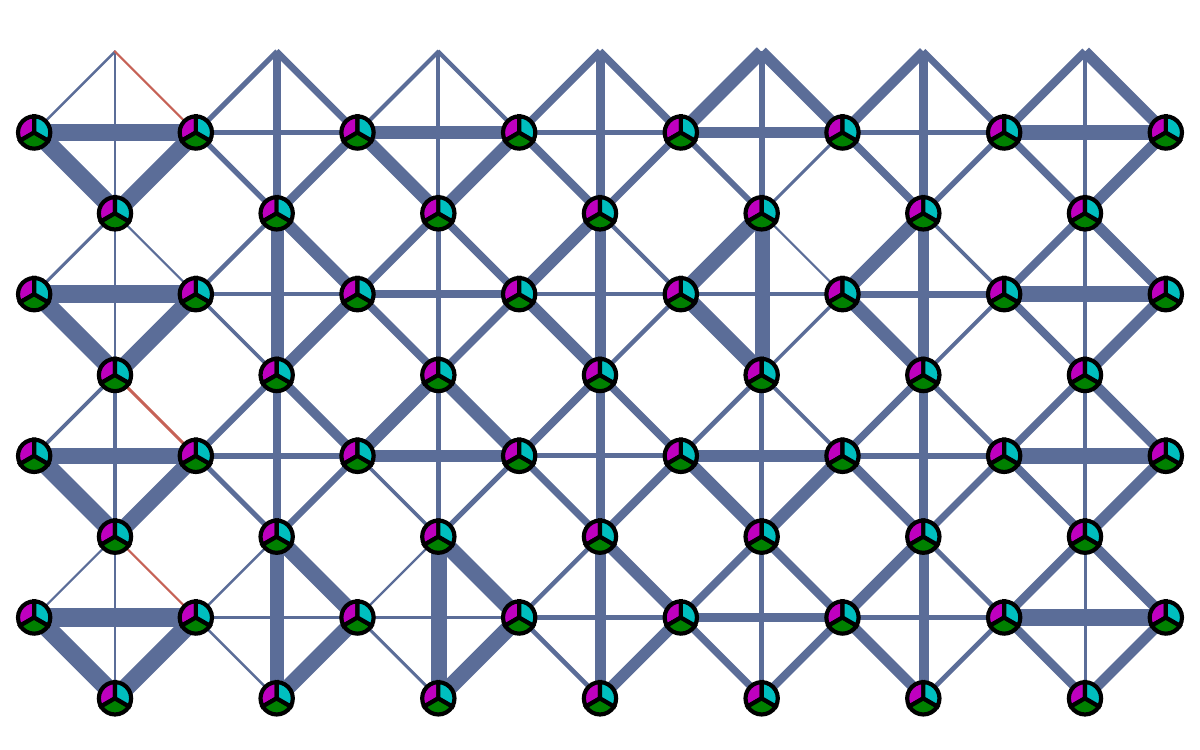}}
  \caption[local_unitcell_4x7]{%
   (\ref{fig:local_unitcell_4x7_a}) Ground state and (\ref{fig:local_unitcell_4x7_b}) local minimum state real-space bond strength obtained by DMRG simulation for the system size of $\left(7,4\right)$. %
    The energy of the states are (\ref{fig:local_unitcell_4x7_a}) $E_0 = -201.698$, (\ref{fig:local_unitcell_4x7_b}) $E_1 = -201.468$. %
    }\label{fig:local_unitcell_4x7}
\end{figure}

After confirming the absence of three-sublattice antiferromagnetic (AFM) and simple triangular valence cluster solid (VCS) orders, we focus on the lattice's unit cell direction.
The density matrix renormalization group (DMRG) simulations converge to several distinct candidate patterns in valence bond strength, all with closely related energies.
Some patterns exhibit apparent local triangular configurations; however, these triangles do not form a clear periodic arrangement characteristic of a VCS phase.
In the system size $\left(7,4\right)$ — which precludes long singlets crossing the boundary and includes a total of 60 sites (a multiple of 3) — the local valence bond strength of the lowest energy state displays a stripe pattern (Fig.~\ref{fig:local_unitcell_4x7_a}).
This stripe pattern has a period of three unit cell lengths in the $ x $-direction and maintains translational symmetry in the $ y $-direction.
% The energy of this local minimum state is also a little higher than the stripe pattern state.

We also calculate the ground state of larger system sizes, up to $(10,6)$ (Details in Supplemental Materials).
There are local minimum states for different system sizes. We summarize some of the lowest energy states for different sizes in Table~ \ref{table:energy_kite}.

\begin{table*}[htp]
    \small
    \begin{center}
        \begin{tabular}{c|cccccc}
            \hline\hline
            Algorithm    & Size                        & Number of Spins    & Number of bonds    & Valence Bond Coupling Structure               & Energy   & Energy per bond  \\
            \hline\hline
            ED    & $\left(3,3\right)$                 & 18                 & 54                 & Nematic(Fig. \ref{fig:ed_pattern_b})& -64.343  & -1.1915 \\
            \hline
            \multirow{17}*{DMRG}
                  & \multirow{2}*{$\left(7,4\right)$}  & \multirow{2}*{60}  & \multirow{2}*{168} & Triangle(Fig. \ref{fig:local_unitcell_4x7_b})                & -201.468 & -1.1992 \\
                  &                                    &                    &                    & Stripe(Fig. \ref{fig:local_unitcell_4x7_a})                & -201.698 & -1.2006 \\
            \cline{2-7}
                  & \multirow{2}*{$\left(10,4\right)$} & \multirow{2}*{84}  & \multirow{2}*{240} & Stripe                                                                                                   & -286.548 & -1.1940 \\
                  &                                    &                    &                    & Triangle \& Stripe                                                                                       & -286.283 & -1.1928 \\
            \cline{2-7}
                  & \multirow{2}*{$\left(7,5\right)$}  & \multirow{2}*{75}  & \multirow{2}*{210} & Incompatible                                                                                             & -250.520 & -1.1929 \\
                  &                                    &                    &                    & Stripe                                                                                                   & -250.928 & -1.1949 \\
            \cline{2-7}
                  & \multirow{2}*{$\left(7,6\right)$}  & \multirow{2}*{90}  & \multirow{2}*{252} & Incompatible                                                                                             & -299.661 & -1.1891 \\
                  &                                    &                    &                    & Stripe                                                                                                   & -300.598 & -1.1928 \\
            \cline{2-7}
                  & $\left(8,6\right)$                 & 102                & 288                & Triangle                                                                                                 & -342.361 & -1.1887 \\
            \cline{2-7}
                  & \multirow{3}*{$\left(9,6\right)$}  & \multirow{3}*{114} & \multirow{3}*{324} & Incompatible                                                                                             & -384.193 & -1.1858 \\
                  &                                    &                    &                    & Triangle                                                                                                 & -384.070 & -1.1854 \\
                  &                                    &                    &                    & Triangle \& Stripe                                                                                       & -384.550 & -1.1868 \\
            \cline{2-7}
                  & \multirow{2}*{$\left(10,6\right)$} & \multirow{2}*{126} & \multirow{2}*{360} & Triangle                                                                                                 & -426.819 & -1.1856 \\
                  &                                    &                    &                    & Stripe                                                                                                   & -426.903 & -1.1858 \\
            \hline
            %iDMRG & $\left(\infty(2),6\right)$         & 24                 & 72                 &                                                    & -84.361  &         \\
        \end{tabular}
    \end{center}
    \caption[ ]{Energy of the ground state and local minimum states for different system sizes. %
    Some of the real space bond strength structures are shown in Supplemental Materials.}
    \label{table:energy_kite}
\end{table*}

We have identified several coupling configurations in this model, particularly near the system boundaries.
For straight open boundaries extending over more than five unit cells, the coupling structure forms repeating triangular patterns, breaking the reflection symmetry along the boundary.
In cases with a four-unit cell width, the reflection-symmetric boundary can be viewed as a superposition of triangular arrays oriented differently.
The stronger coupling strength on the bonds perpendicular to the boundary suggests that these triangular singlets connect to form a boundary ring.
Applying a reflection transformation to this boundary ring does not alter the system's energy.

For systems with length $3n + 1$ in the $x$ direction, the stripe phase with a 3-unit-cell stripe exhibits lower energy.
This pattern comprises a one-unit-cell-wide triangular array at the boundary and a two-unit-cell-wide reflection-symmetric structure.
We refer to the strong coupling bonds as flat stripes.
The energy associated with these flat stripes is lower than that of two corresponding staggered triangular VCS patterns.
By comparing the low-energy states of systems with identical sizes, we observe that configurations with a greater number of stripe patterns tend to have lower energies.

For systems with other lengths in the $x$ direction, the stripe phase cannot fit the system well.
When the length in the $y$ direction is 6, a clear triangular pattern emerges.
This pattern exhibits a 2-unit-cell-wide periodic triangular configuration away from the open boundary.
As the length in the $x$ direction increases, this triangular configuration forms a vertical stripe-like arrangement.
Consequently, we deduce that the stripe pattern reduces the local energy, while the boundary pattern constrains the global coupling structure.

\begin{figure}[htbp]
    \centering
    \includegraphics[width=0.45\textwidth]{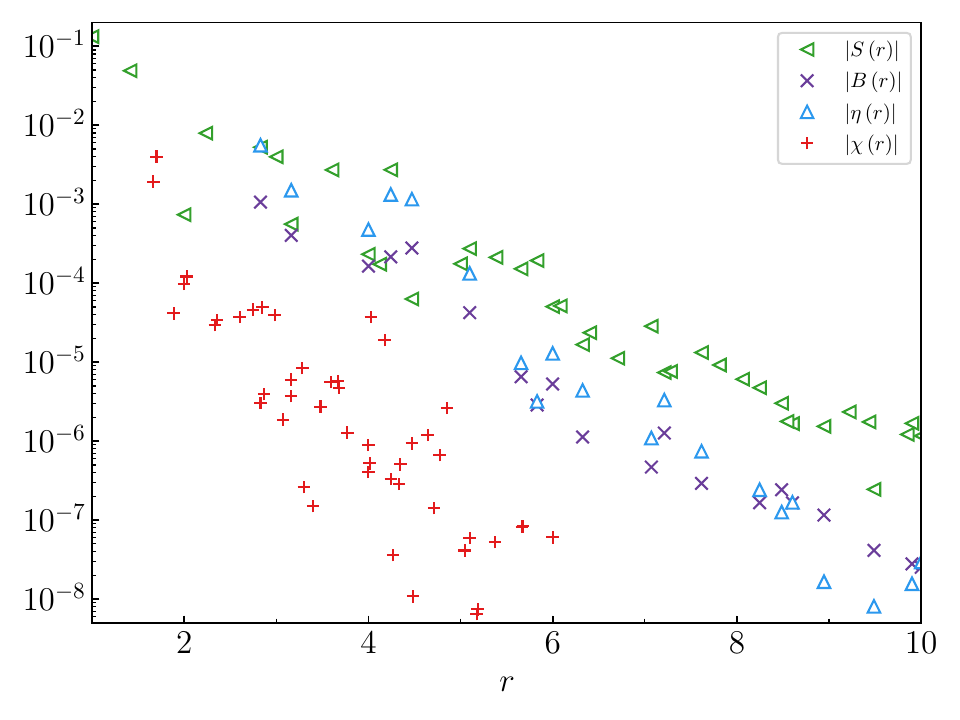}
    \caption[multiple correlation]{%
        Spin correlation $S\left(r\right)$, bond correlation $B\left(r\right)$, nematic correlation $\eta\left(r\right)$ and chiral correlation $\chi\left({r}\right)$ measured in system with size $\left(N_x,N_y\right)=\left(10,6\right)$. %
        They all show exponential decay. The correlation lengths are $ {r_S} = 0.89 a $, $r_{B} = 0.55 a$, $r_{\eta} = 0.61 a$ and $r_{\chi} = 0.32 a$, $a$ is the length of square side. %
    }\label{fig:kite_corre}
\end{figure}

For the stripe phase, we calculate several correlation functions and show them in Fig. ~\ref{fig:kite_corre}.
It includes spin correlation function Eq.~(\ref{eq:spin_correlation}), chiral correlation function
\begin{equation}\label{eq:chiral_correlation}
  \chi\left({r}\right)=\left<\chi_{ijk} \chi_{i'j'k'}\right>,
\end{equation}
and nematic correlation function
\begin{equation}
  \eta\left({r}\right)=\left<\eta_{x} \eta_{x+r}\right>-\left<\eta_{x}\right>\left< \eta_{x+r}\right>.
\end{equation}
The local chiral and nematic operators are defined as $\chi_{ijk}=i\left(P_{ijk}-P^{-1}_{ijk}\right)$ and $\eta_{\text{i}}=\left(B_{\text{i}}^{x}-B_{\text{i}}^{y}\right)$, where $P_{ijk} = P_{ij}P_{jk}$ and $B_{i}^{x,y}$ represents the bond coupling along the $x$ and $y$ directions within a unit cell.
These correlation functions exhibit exponential decay, with a short correlation length less than the side length of the square.
This result excludes the possibility of chiral and nematic orders in the checkerboard lattice.

%%%
\section{bond stripe pattern}\label{sec:stripe_pattern}

Bond stripe patterns have been identified in mean-field calculations for the $\mathrm{SU}(N)$ Heisenberg model using self-consistent minimization algorithms.
These patterns emerge as low-energy states for $N=4,5$ on the triangular lattice and $N=5,6$ on the square and $J_1$-$J_2$ honeycomb lattices~\cite{hermele_mott_2009,yao_topological_2021,yao_intertwining_2022}.
However, in contrast to the doubled unit-cell bond stripe states found in these studies, the stripe phase observed in our work exhibits a tripled unit-cell structure.
Additionally, while the previously discovered stripe states possess $U(1)$ gauge flux, this characteristic is absent in our findings.
Moreover, mean-field calculations have shown that for $N=3$, the ground state is a VCS state, comprising singlets formed by 3 or 6 sites.
It is important to note that the applicability of the large-$N$ approximation employed in the mean-field approach may be limited.
Our computations indicate that the $J_\times$ interaction suppresses both the 3-sublattice order and the global VCS order, leading to a bond stripe phase distinct from those predicted by mean-field results.

\begin{figure}[htbp]
    \centering
        \includegraphics[width=0.45\textwidth]{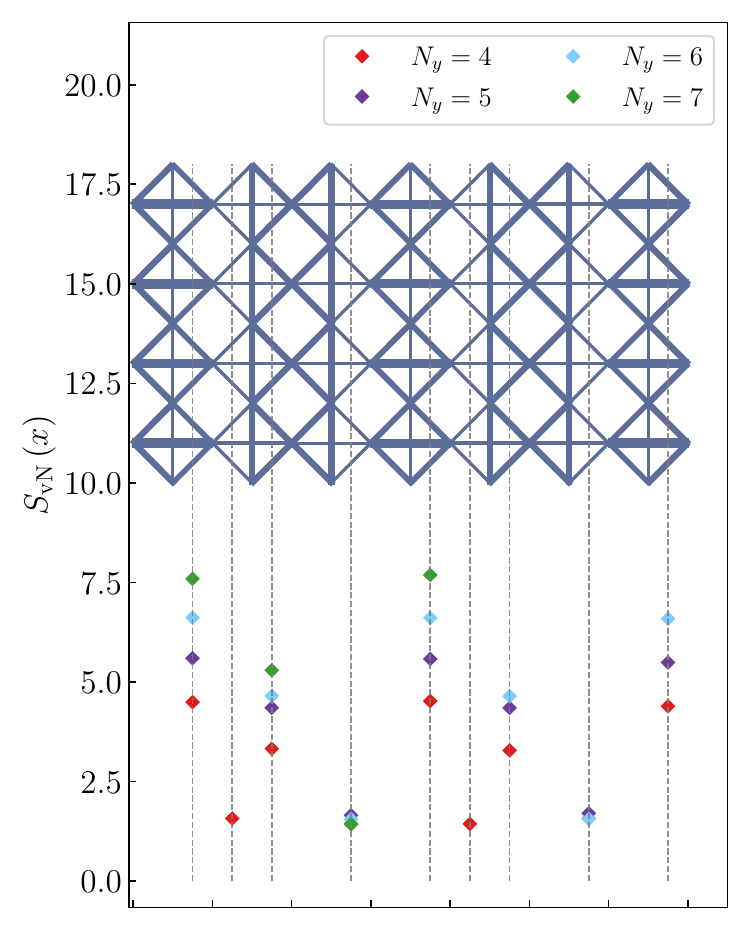}
    \caption[entanglement entropy]{%
        Entanglement entropy for the division at different positions. %
        The dashed lines mark the system division position, and the red bars below represent the value of entanglement entropy at the dashed line. %
        It shows distinct periodicity, with approximate minima between stripes and maxima on cutting triangle arrays. %
    }\label{fig:ee_xx7_cut}
\end{figure}

The flat regions of the stripe pattern can break translational symmetry, transitioning into a triangular valence cluster solid (VCS) pattern when interfacing with boundary triangular arrays or stripe corners.
This suggests that the state is a direct product state.
We compute the entanglement entropy for subsystems partitioned along various lines, as depicted in Fig.~\ref{fig:ee_xx7_cut}.
Additionally, we analyze the dependence of entanglement entropy on the edge length between stripes.
Our findings reveal that the entanglement entropy for subsystems intersecting weak bonds is minimal and remains approximately constant regardless of the subsystem boundary lengths.
The periodicity observed in the entanglement entropy further indicates that entanglement is confined to a minimal range.
Within the stripes, spins form singlets on the occupied sites, explaining the inherently short-range correlations across the stripes.

\subsection*{Decoupled Stripe and Single-Stripe Ladder}
Since each stripe is nearly pure, it is feasible to compute the correlation within a stripe on the ladder stripe that it occupies.
We use ED to reconstruct the single-stripe patterns in a system comprising four to six unit cells.

We obtain results for the triangular array comparable to the previous exact solution for the spin tetrahedron chain~\cite{chen_exact_2006}.
The ground state exhibits two-fold degeneracy, which can be lifted by breaking the reflection symmetry. % or adjusting the strength of the NNN interaction in the direction of the unit cell base vector.
The symmetry-broken states exhibit patterns similar to those of triangular arrays in stripe states.
These states are pure direct product states.
The flat stripe pattern is also reproduced in isolated ladder systems.
In these systems, the ground state is unique and features uniform coupling strength $B=-0.1986$ on each bond.

\begin{figure}[htbp]
    \centering
    \subfloat[Single triangular array ladder. \label{fig:kite_chain_1}]{
        \includegraphics[width=0.40\textwidth]{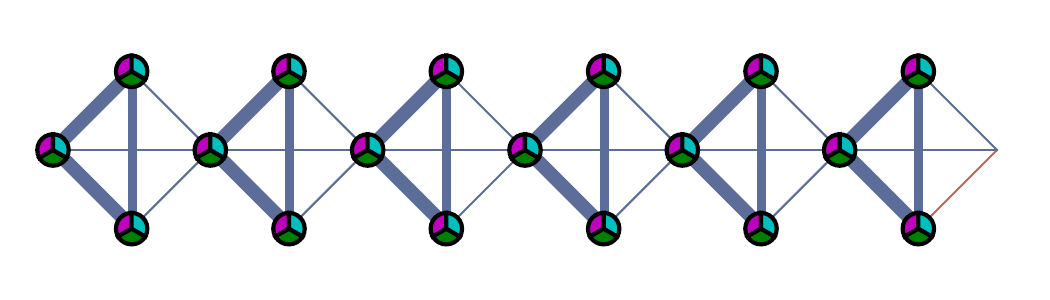}}
    \hspace{0.05\textwidth}
    \subfloat[Flat stripe ladder. \label{fig:plain_chain_1}]{
        \includegraphics[width=0.40\textwidth]{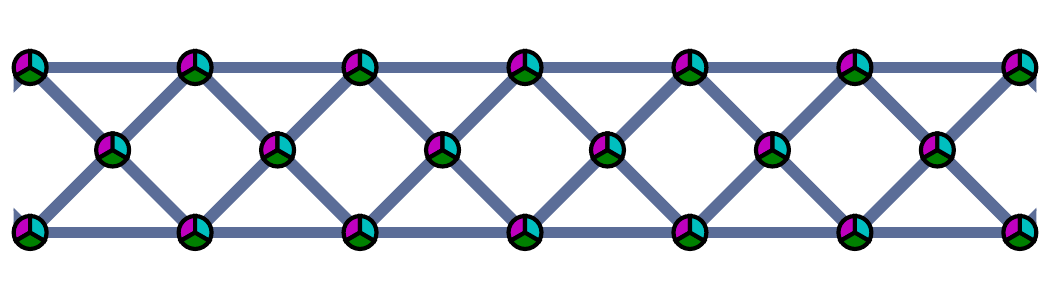}}\\
    %\hspace{0.05\textwidth}
    \subfloat[Correlation function in flat stripe ladder. \label{fig:plain_chain_corre}]{
        \includegraphics[width=0.45\textwidth]{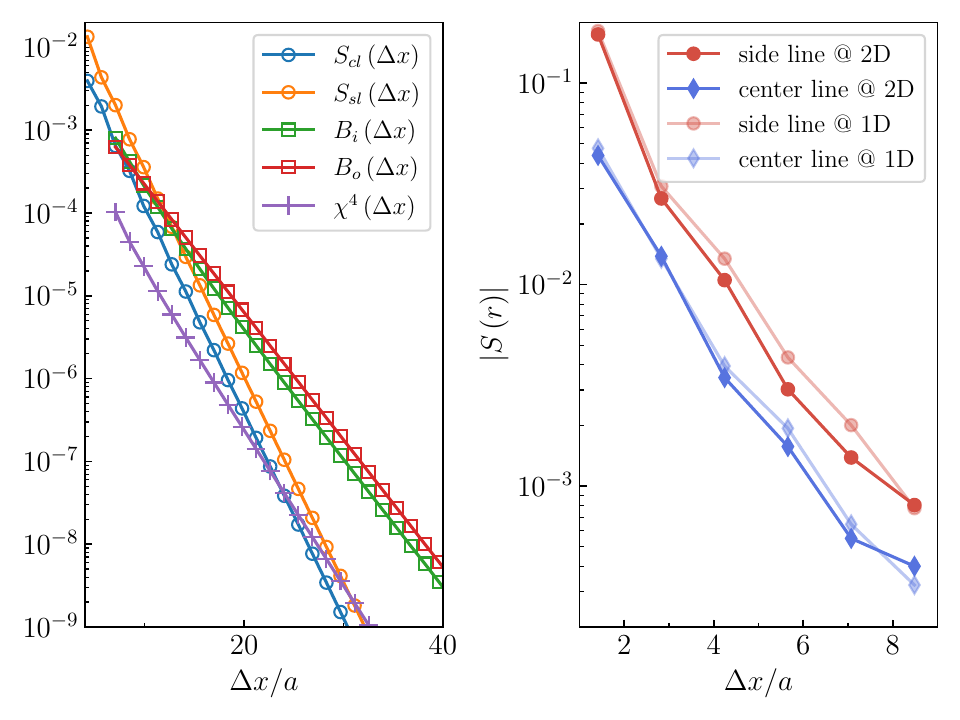}}
    \caption[1]{%
        $\left(a\right)$,$\left(b\right)$ the local spin and bond pattern of the triangle array ladder and the flat stripe ladder, respectively.
        $\left(c\right)$ (left) Spin correlation along center line $S_{cl}$ and side line $S_{sl}$.
        Bond correlation of inner bonds $B_{i}$ and outer bonds $B_{o}$.
        (right) Comparison of spin correlation functions for flat stripes in 2D and 1D systems.}%
    \label{fig:single_ladder}
\end{figure}

Building upon the successful reproduction of two stripe patterns in isolated ladder systems using ED, we further employ the DMRG algorithm to investigate long-range correlation functions within the stripes.
The real-space pattern is depicted in FIG.~\ref{fig:kite_chain_1}.
The results for the triangular array stripe ladder align with those from ED calculations, indicating a direct product of triangular singlets whose correlation functions correspond to the VCS state.

For the flat stripe ladder system, we compute the system length up to 64 unit cells, maintaining a maximum kept dimension $D = 1467$ to ensure truncation errors remain below $10^{-10}$.
This limited bond dimension suggests that this quasi-one-dimensional system is gapped.
As shown in FIG.~\ref{fig:plain_chain_1}, the DMRG calculations reveal no spontaneous breaking of translational or spatial inversion symmetry.
Examining the valence bond strengths away from the boundary, we observe differences between bonds parallel to the chain direction ($B_{=}=-0.1848$) and those diagonal to it ($B_{\times}=-0.1993$).
This is similar to results for the two-dimensional case, where the ratio of these bond strengths is approximately 0.93 (e.g., $B^{2d}_{=}=-0.175$ and $B^{2d}_{\times}=-0.1846$ in the $\left(10,6\right)$ system).

The correlation functions of the flat stripe ladder system closely resemble those of the two-dimensional case.
The spin and bond correlation functions for this system are presented in FIG.~\ref{fig:plain_chain_corre}.
Both exhibit exponential decay with correlation lengths $r_{S}^{1d} = 1.73a$ and $r_{B}^{1d} = 2.82a$, respectively.
We calculate the corresponding four-point chirality correlation function
\begin{equation}
    \chi^{4}_{ijkl}=\frac{i}{4}\left(P_{ijkl}-P^{-1}_{ijkl}\right),
\end{equation}
where $i, j, k, l$ are defined on the square plaquette.
These functions also decay exponentially, failing to corroborate the gauge flux of the stripe state observed in mean-field calculations.

\section{Summary and Discussion}\label{sec:sum}

The model exhibits frustration when three-site singlets are placed on vertex-connected tetrahedra forming pyrochlore lattices.
Using ED and DMRG simulations, we study the SU(3) spin model on the checkerboard lattice, also known as the planar pyrochlore.
By tuning the diagonal couplings, we observe a quantum phase transition between the square lattice model and the checkerboard lattice model.
The addition of diagonal couplings suppresses the 3-sublattice AFM order.
Stripe patterns frequently emerge in DMRG calculations, and the stripe state serves as the global ground state for suitable system sizes.
The twofold degeneracy of the ground state observed in ED calculations is consistent with the $C_{4}$ symmetry breaking of the stripe phase.
In this phase, the stripes decouple, leading to short-range correlation functions.
The stripe patterns consist of a triangular array and a uniform flat stripe.
They can be reproduced in a ladder model defined on their sites, where correlations within the stripes remain short-ranged.

In previous mean-field studies of the SU(3) spin model on various lattices~\cite{yao_topological_2021,yao_intertwining_2022}, the lowest energy state is the VCS.
For larger $N$, stripe and chiral spin liquid phases have been discovered.
The VCS-stripe mixed state identified here can be viewed as a step closer to the chiral spin liquid phase, according to general phase diagram results~\cite{boos_time-reversal_2020,hermele_mott_2009,yao_topological_2021}.
Further investigation is required to determine whether additional interactions can induce a spin liquid phase.

Furthermore, under the zeroth-order approximation, the ground states of the model consist of local singlets on three of the vertices of a planar tetrahedron.
These degenerate ground states can adopt different triangular configurations.
The various triangle patterns span a configuration space analogous to a quantum trimer model (details in Supplemental Materials)~\cite{jandura_quantum_2020}.
An arrow representation can be constructed in line with studies of the quantum dimer model on the kagome lattice~\cite{misguich_quantum_2002}.
Since each site is shared by two tetrahedra and is associated with a singlet, we can define an arrow on each site pointing toward the tetrahedron to which the singlet belongs.
This arrow representation forms a restricted vertex model, potentially related to $\mathbb{Z}_2$ gauge theory~\cite{yan_classical_2024}.

%%%%%%%%%%%%%%%%%%
\section{Acknowledgments}
This work was supported by the National Key Research and Development Program of China Grant No. 2022YFA1404204 and the National Natural Science Foundation of China (Grants No. 11625416 and No. 12274086).
The authors acknowledge Fudan High Performance Computation Center for providing HPC resources that have contributed to the numerical results reported in this paper.

%%%%%%%%%%%%%%%%. References
\bibliography{checkerboardsu3}

\pagebreak

\newpage
\widetext
\begin{center}
\textbf{\large Supplemental Materials for ``Bond Stripe Patterns in $\rm{SU}\left(3\right)$ spin model on checkerboard lattice"}
\end{center}
%\begin{center}

%\end{center}

\setcounter{section}{0}
\setcounter{equation}{0}
\setcounter{figure}{0}
\setcounter{table}{0}
\makeatletter
\renewcommand{\thefigure}{S\arabic{figure}}
\renewcommand{\thetable}{S\arabic{table}}
\renewcommand{\theequation}{S\arabic{equation}}
\renewcommand{\bibnumfmt}[1]{[S#1]}
\renewcommand{\citenumfont}[1]{S#1}
\renewcommand{\thesection}{S\arabic{section}}
\makeatother

\section{DMRG simulation in square frame}\label{sm:dmrg_sq}
\begin{figure}[htbp]
    \centering
    \subfloat{\includegraphics[width=0.4\textwidth]{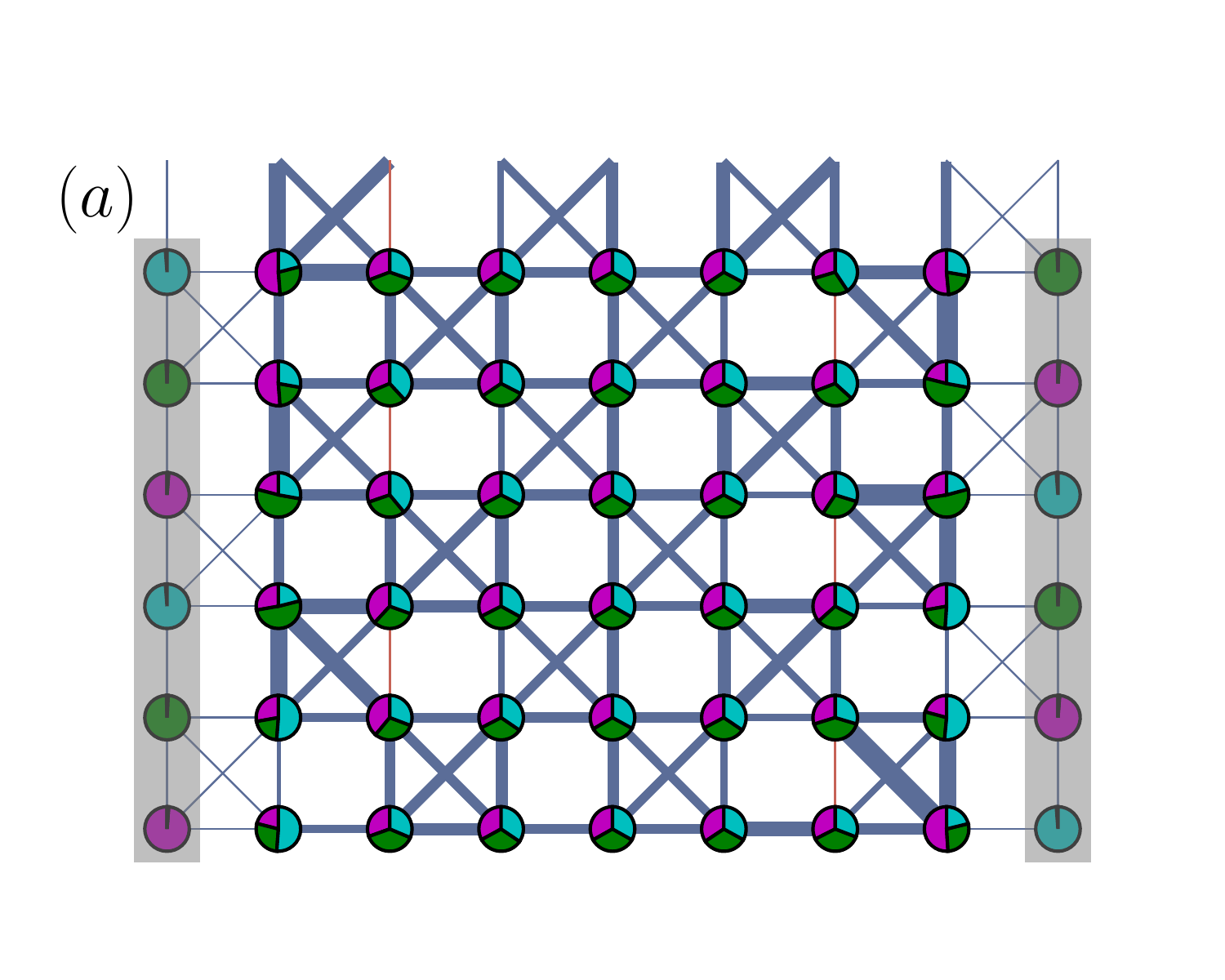}}
    \hspace{0.05\textwidth}
    \subfloat{\includegraphics[width=0.4\textwidth]{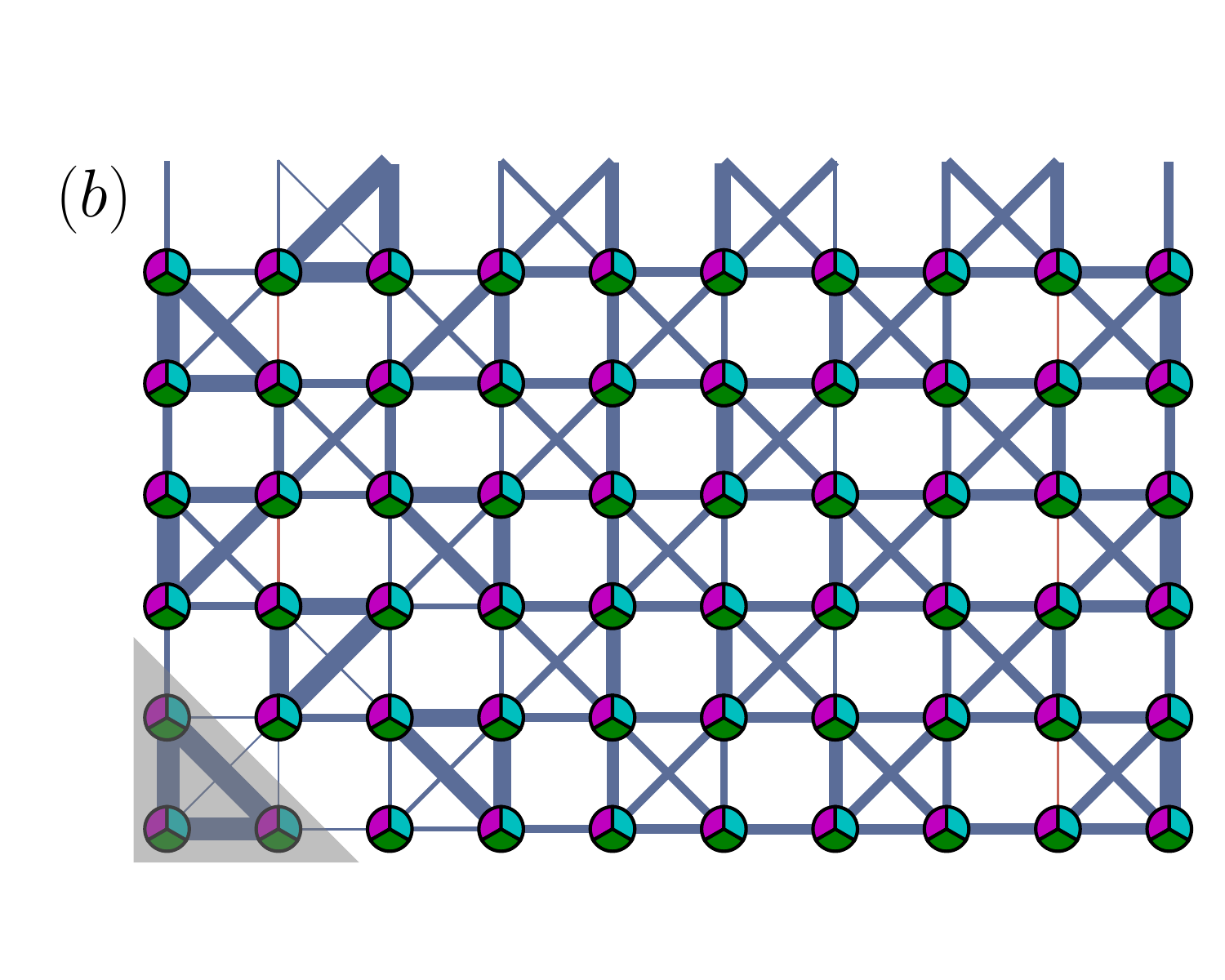}}
    \hspace{0.05\textwidth}
    \subfloat{\includegraphics[width=0.45\textwidth]{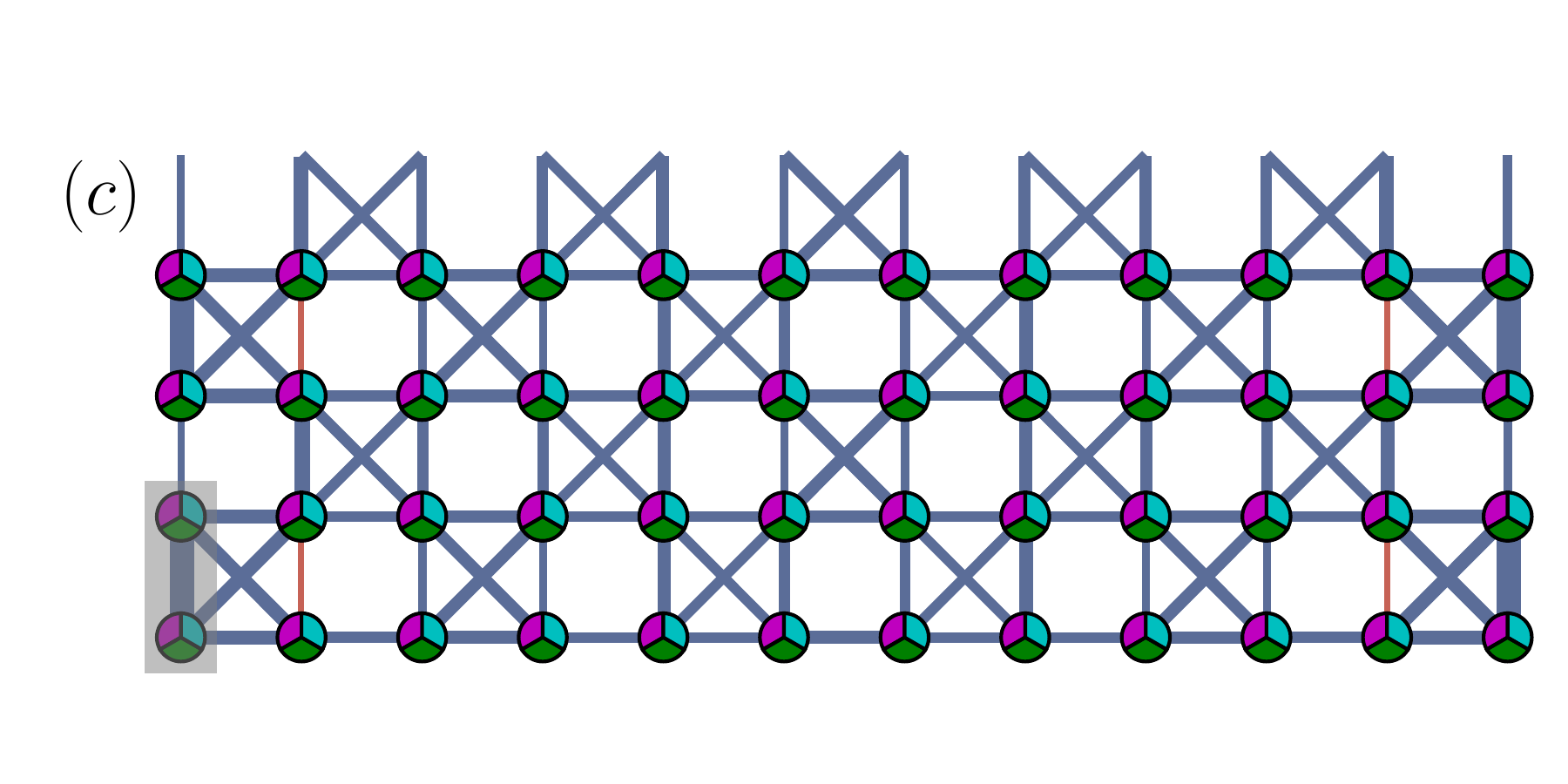}}
    \hspace{0.05\textwidth}
    \subfloat{\includegraphics[width=0.45\textwidth]{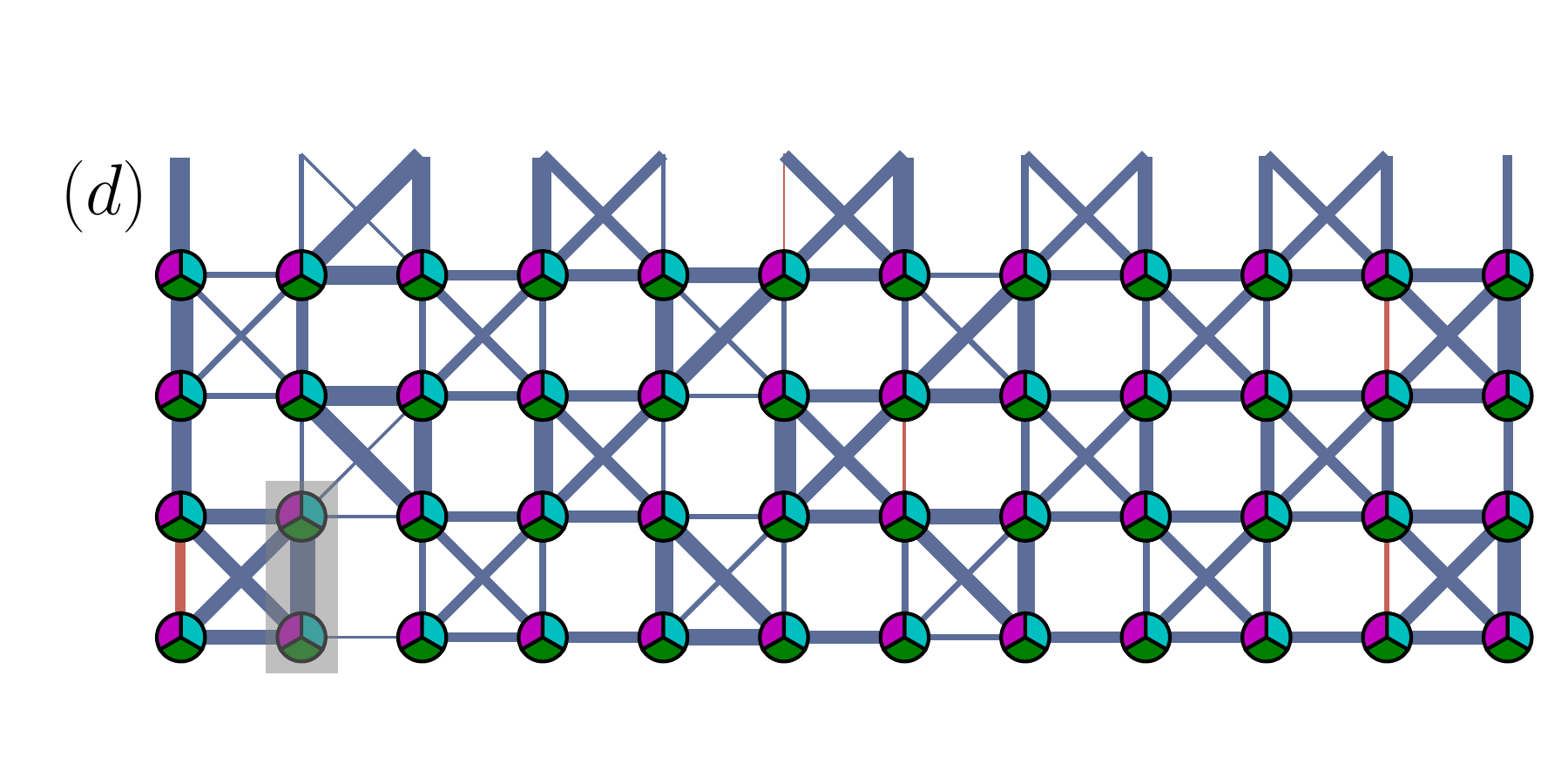}}
    \captionsetup{justification=raggedright,singlelinecheck=false}
    \caption[induction on square lattice]{%
        Bond energies and local color densities on a square lattice with the different inducing structures on the boundary (gray-covered region). %
        Open boundary condition on $x$ direction and periodic boundary condition on $y$ direction. %
        $\left(a\right)$ External fields inducing AFM order applied on the boundary sites; %
        $\left(b\right)$ Stronger interaction on bonds in a triangle; %
        $\left(c\right)$ Stronger interaction on a single bond; %
        $\left(d\right)$ Stronger interaction on a single bond with weak correlation in previous results. %
    }\label{fig:field_induce_square}
\end{figure}

The interaction we tune transforms the lattice from a square configuration to a planar pyrochlore structure.
We calculate the ground state patterns along the square lattice directions as a reference to verify the existence of possible orders.
The 3-sublattice AFM order is suppressed, as indicated by the very short penetration depth of the spin polarization (FIG.~\ref{fig:field_induce_square}$\left(a\right)$).
We also attempt to induce valence bond order by applying stronger interactions near the boundary (FIG.~\ref{fig:field_induce_square}$\left(b\right)-\left(d\right)$).
The inhomogeneity of the bonds extends over several unit cells, with staggered bonds induced by the open boundary condition in the $x$ direction remaining.

\section{DMRG simulation of different sizes}\label{sm:dmrg_size}

\begin{figure}[htbp]
  \centering
  \subfloat[$\left(10,4\right)$ Stripe Structure\label{fig:local_unitcell_4x10_a}]{
      \includegraphics[width=0.45\textwidth]{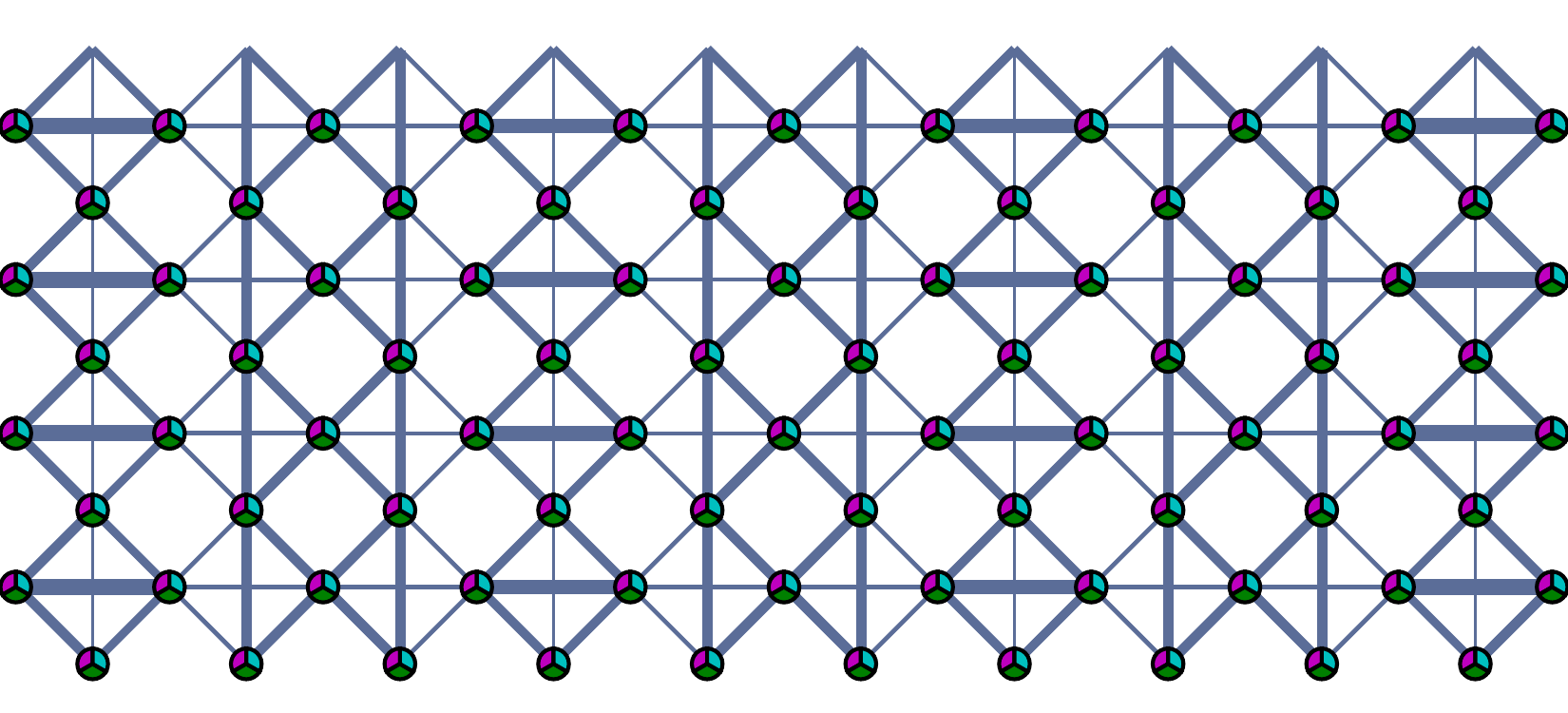}}
  \hspace{0.05\textwidth}
  \subfloat[$\left(10,4\right)$ Local Minimum\label{fig:local_unitcell_4x10_b}]{
      \includegraphics[width=0.45\textwidth]{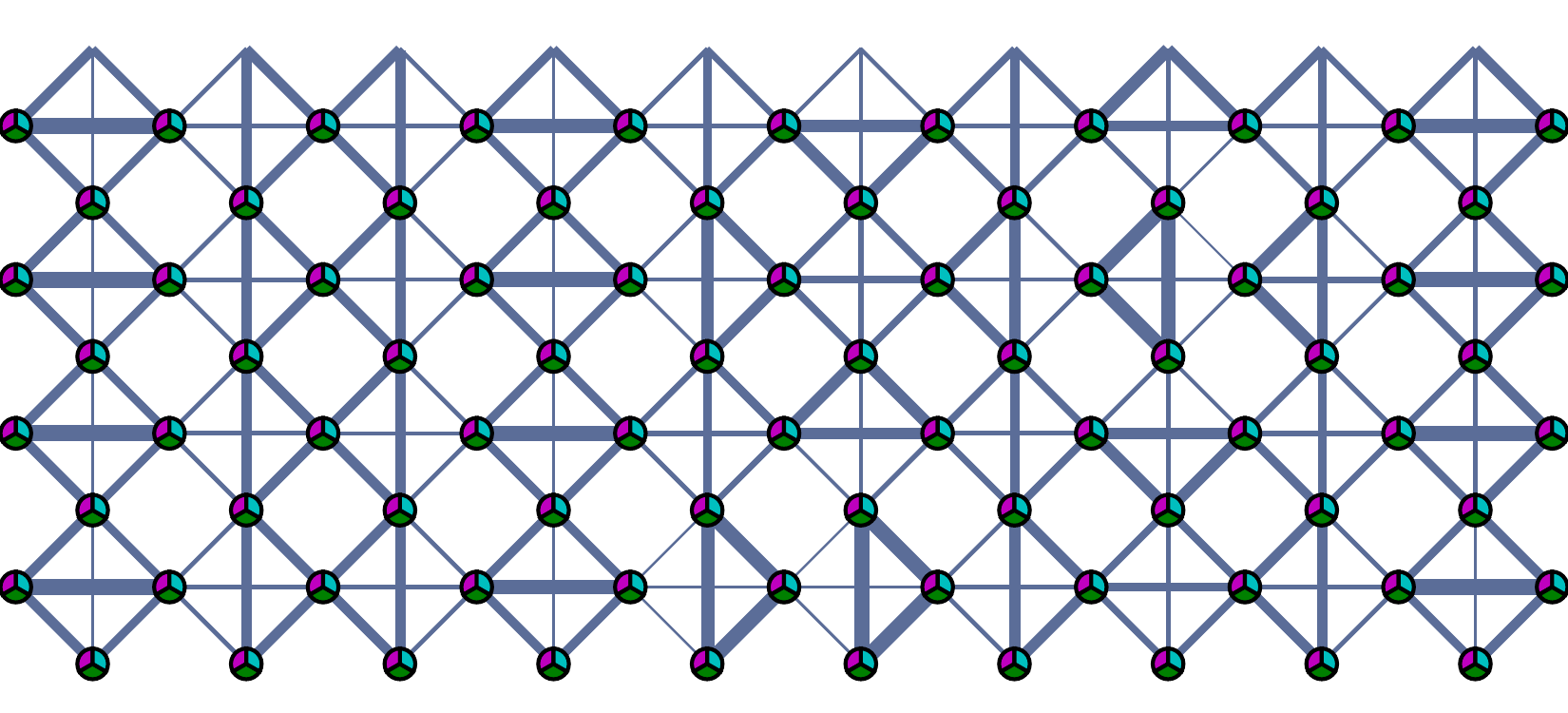}}
\captionsetup{justification=raggedright,singlelinecheck=false}
\caption[local_unitcell_4x10]{%
        Ground state (\ref{fig:local_unitcell_4x10_a}) and local minimum state (\ref{fig:local_unitcell_4x10_b}) real-space bond strength obtained by DMRG simulation for the system size of $\left(10,4\right)$. %
      The energy of the states are (\ref{fig:local_unitcell_4x10_a}) $E_0 = -286.548$, (\ref{fig:local_unitcell_4x10_b}) $E_1 = -286.283$. %
      Their pattern could be regarded as patterns of $7\times 4$ low energy states with an extra stripe period.%
  }\label{fig:local_unitcell_4x10}
\end{figure}

Bond energies and local color densities are presented for different system sizes using DMRG calculations.
We extend the $\left(7,4\right)$ system to $\left(10,4\right)$, ensuring the total number of spins is a multiple of 3, and plot the real-space patterns of the lowest energy state and some competing states in FIG.~\ref{fig:local_unitcell_4x10}.

\begin{figure*}[htbp]
  \centering
  \subfloat[$\left(7,5\right)$ Stripe Structure\label{fig:local_unitcell_yx7_a}]{
      \includegraphics[width=0.35\textwidth]{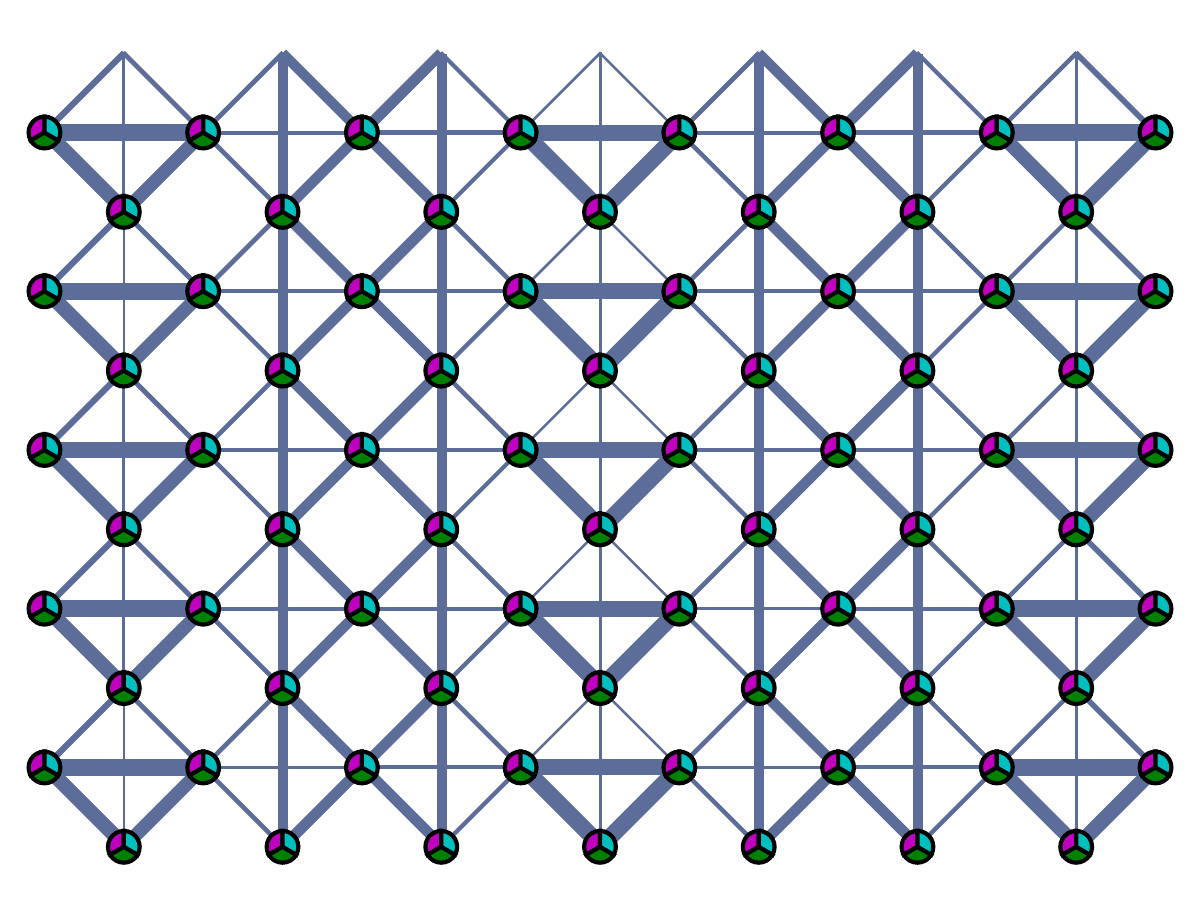}}
  \hspace{0.05\textwidth}
  \subfloat[$\left(7,5\right)$ Local Minimum\label{fig:local_unitcell_yx7_b}]{
      \includegraphics[width=0.35\textwidth]{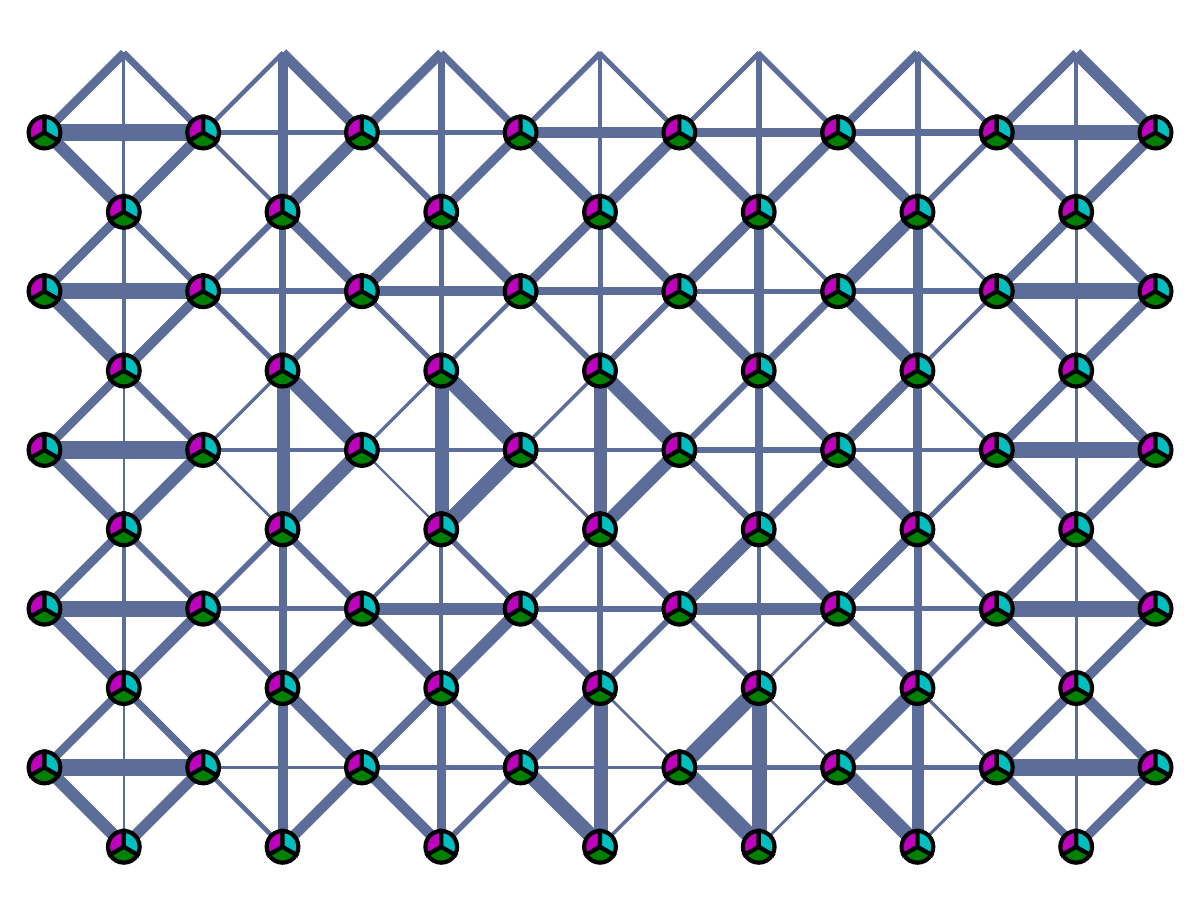}}
  \\
  \subfloat[$\left(7,6\right)$ Stripe Structure\label{fig:local_unitcell_yx7_c}]{
      \includegraphics[width=0.35\textwidth]{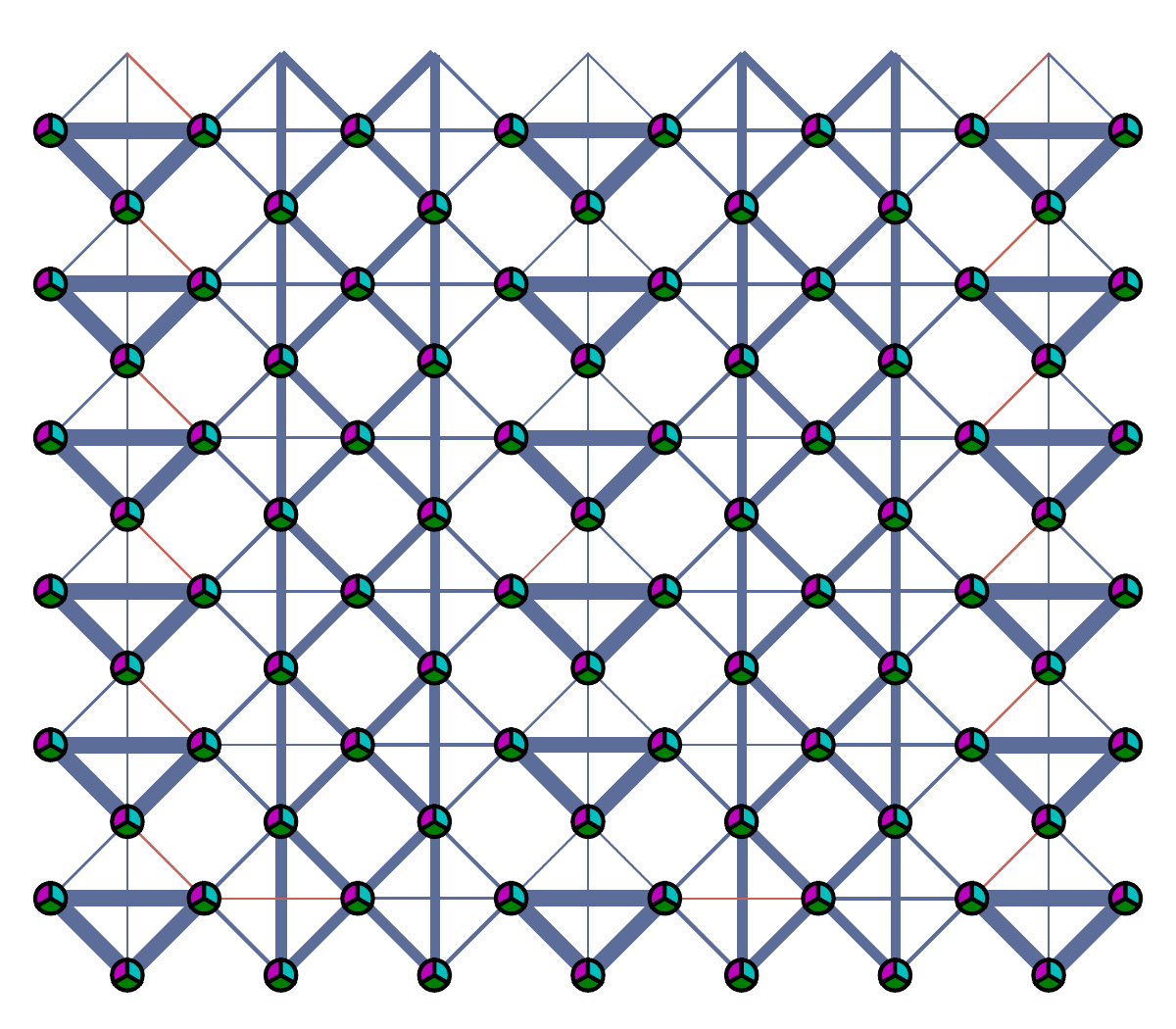}}
  \hspace{0.05\textwidth}
  \subfloat[$\left(7,6\right)$ Local Minimum\label{fig:local_unitcell_yx7_d}]{
      \includegraphics[width=0.35\textwidth]{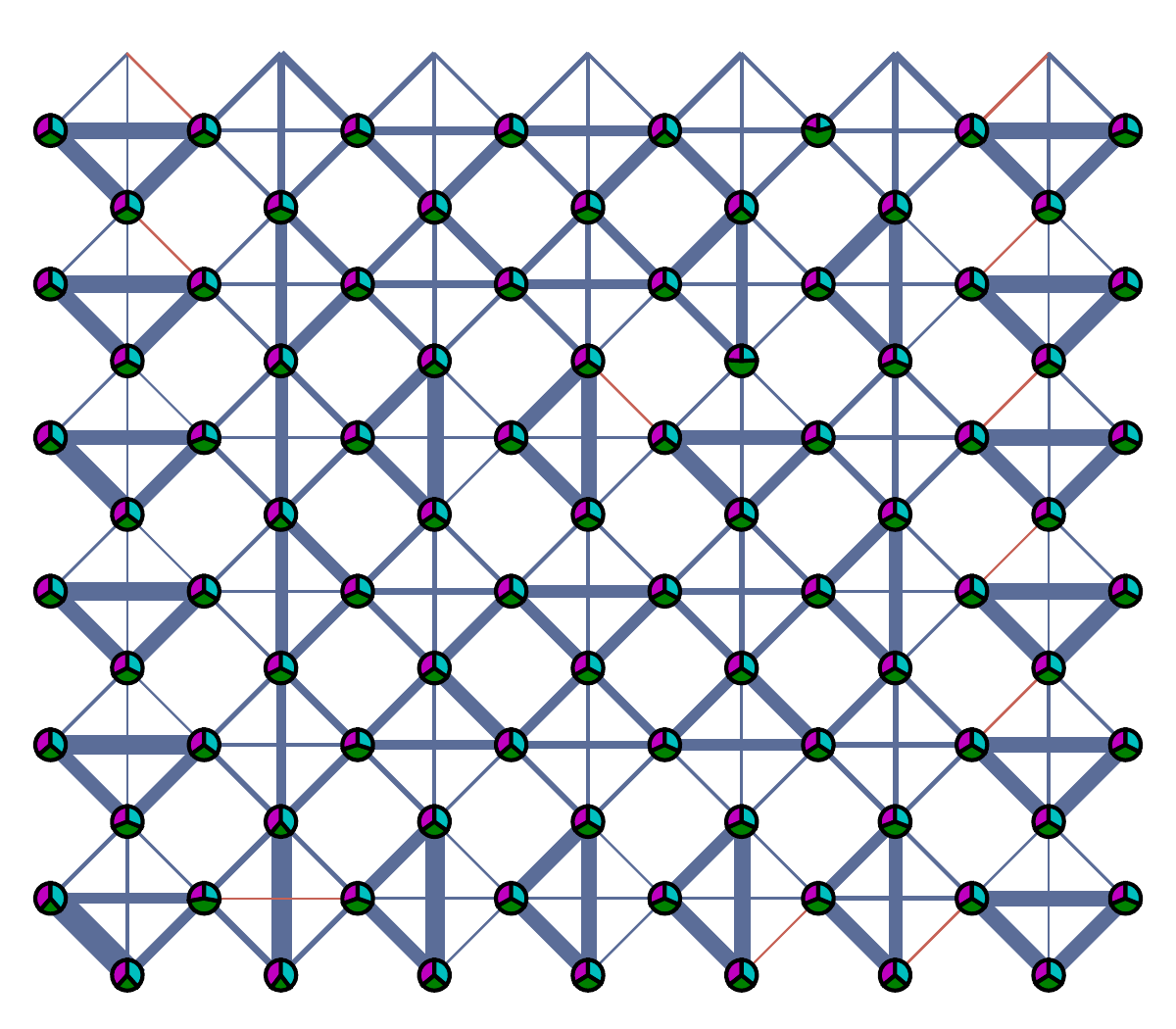}}
      \captionsetup{justification=raggedright,singlelinecheck=false}
  \caption[local_unitcell_yx7]{%
        Ground state (\ref{fig:local_unitcell_yx7_a}, \ref{fig:local_unitcell_yx7_c}) and local minimum state (\ref{fig:local_unitcell_yx7_b}, \ref{fig:local_unitcell_yx7_d}) real-space bond strength obtained by DMRG simulation for the system size of $\left(7,5\right)$ and $\left(7,6\right)$. %
      The energy of the states is (\ref{fig:local_unitcell_yx7_a}) $E_0 = -250.928$, (\ref{fig:local_unitcell_yx7_b}) $E_0 = -250.520$, (\ref{fig:local_unitcell_yx7_c}) $E_0 = -300.598$, (\ref{fig:local_unitcell_yx7_d}) $E_0 = -299.661$. %
      The energy of bond stripe states is lower. %
  }\label{fig:local_unitcell_yx7}
\end{figure*}

Then we extend the system size in the $y$ direction to $\left(7,5\right)$ and $\left(7,6\right)$.
The results are shown in FIG.~\ref{fig:local_unitcell_yx7}.

\begin{figure}[h]
  \centering
  \subfloat[$\left(8,6\right)$ triangle pattern A\label{fig:local_unitcell_6xx_a}]{
      \includegraphics[width=0.36\textwidth]{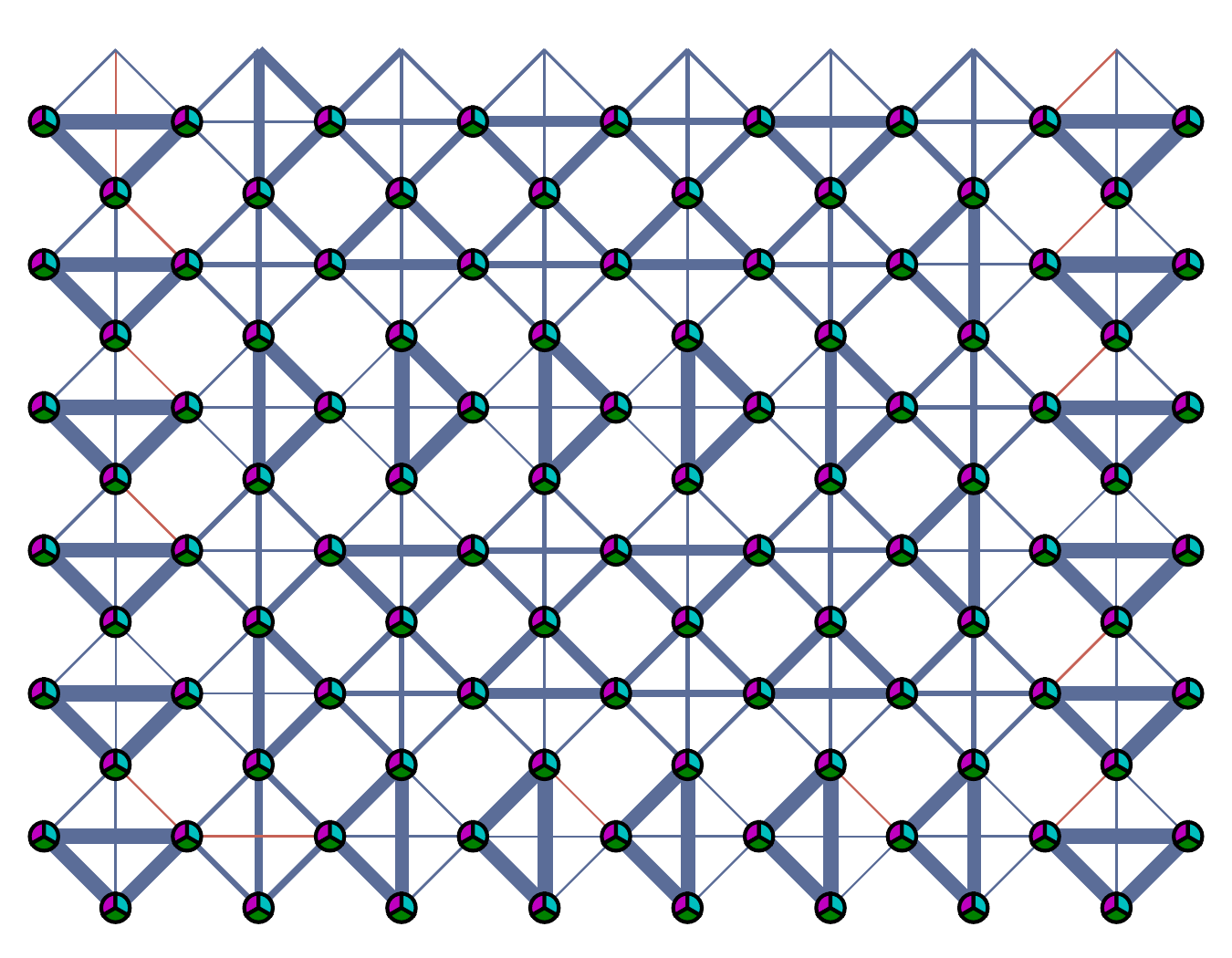}}
  \hspace{0.10\textwidth}
  \subfloat[$\left(9,6\right)$ local minimum\label{fig:local_unitcell_6xx_b}]{
      \includegraphics[width=0.40\textwidth]{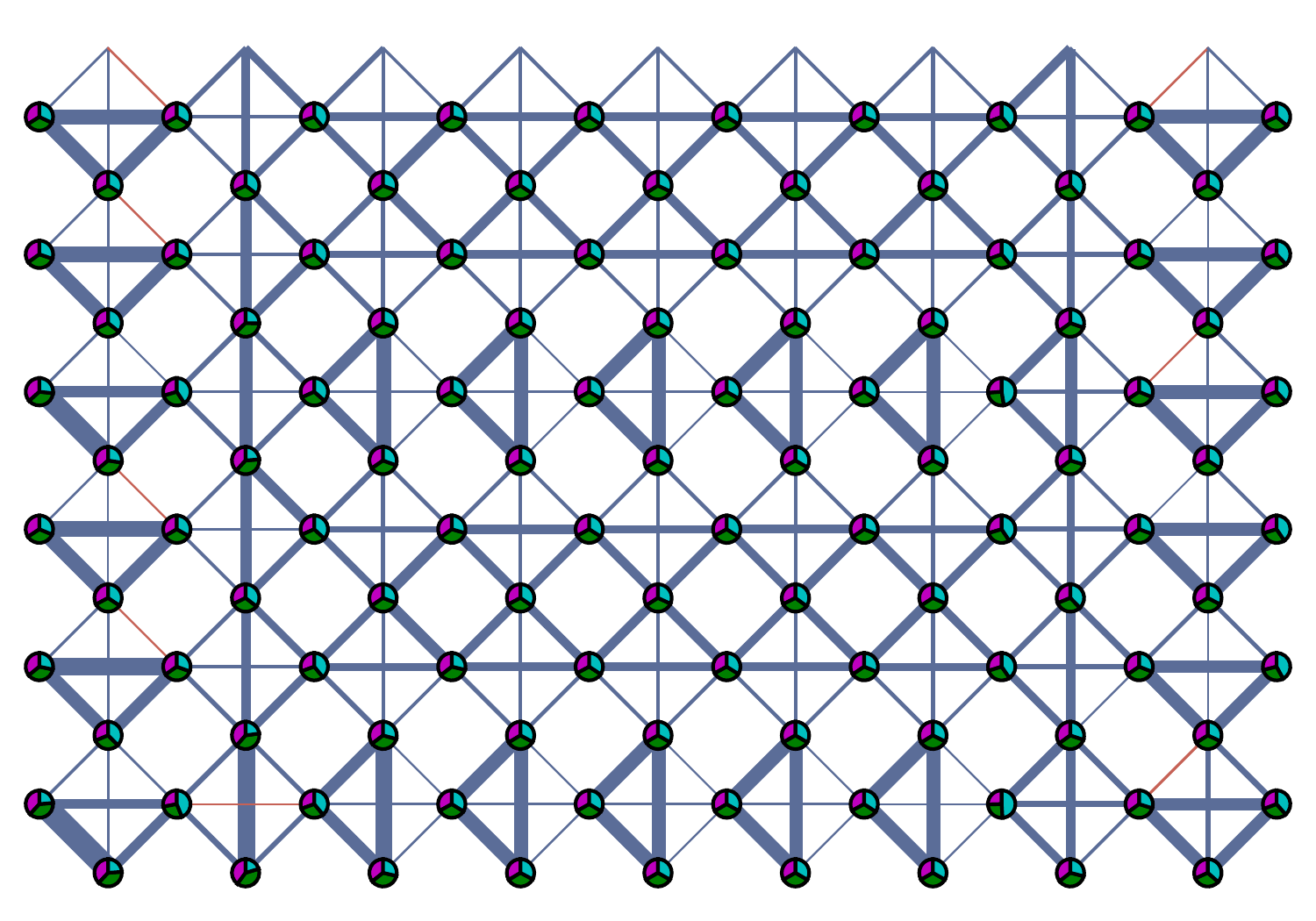}}
  \\
  \subfloat[$\left(9,6\right)$ stripe with triangle pattern A\label{fig:local_unitcell_6xx_c}]{
      \includegraphics[width=0.40\textwidth]{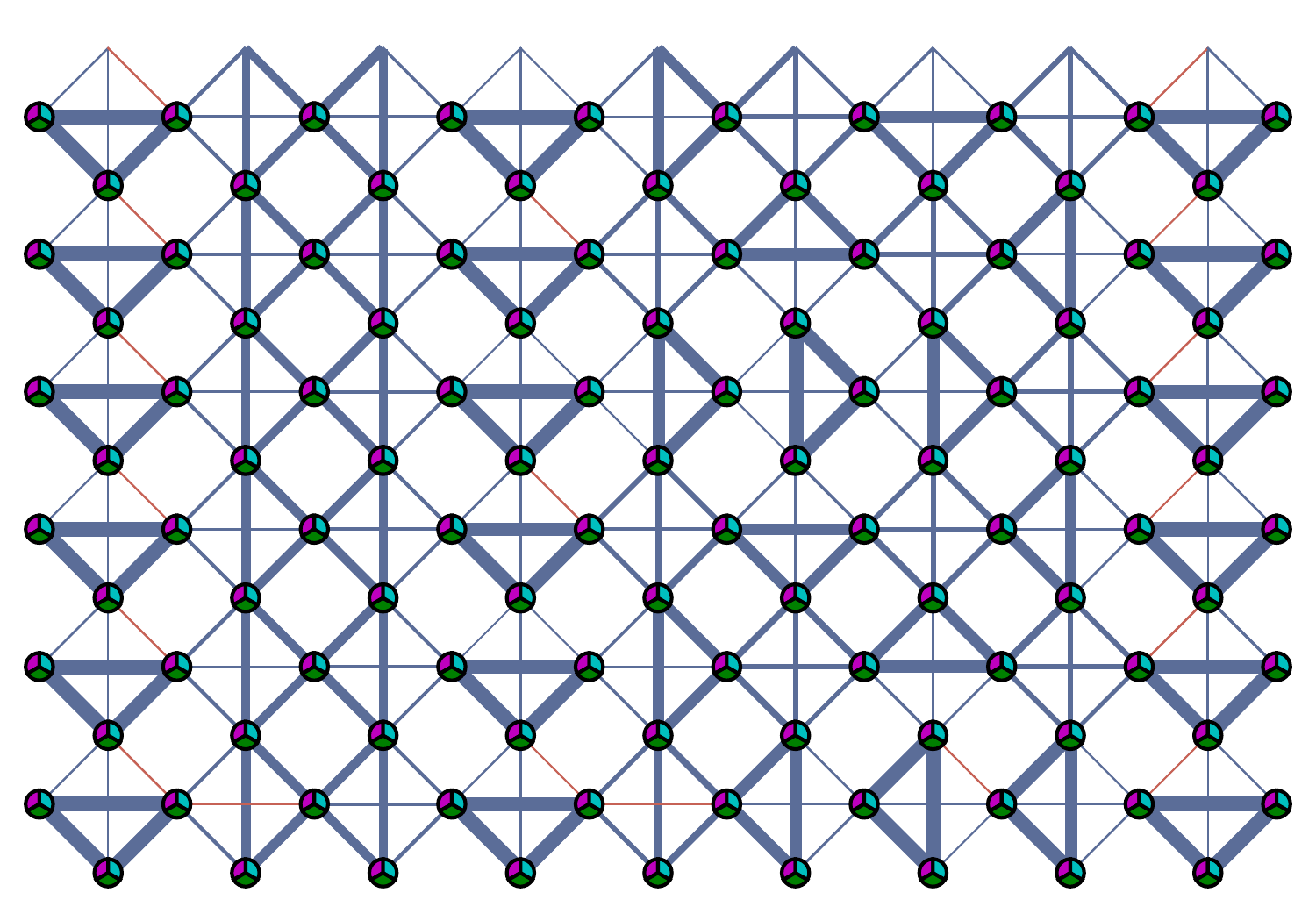}}
  \hspace{0.05\textwidth}
  \subfloat[$\left(9,6\right)$ triangle pattern B\label{fig:local_unitcell_6xx_d}]{
      \includegraphics[width=0.40\textwidth]{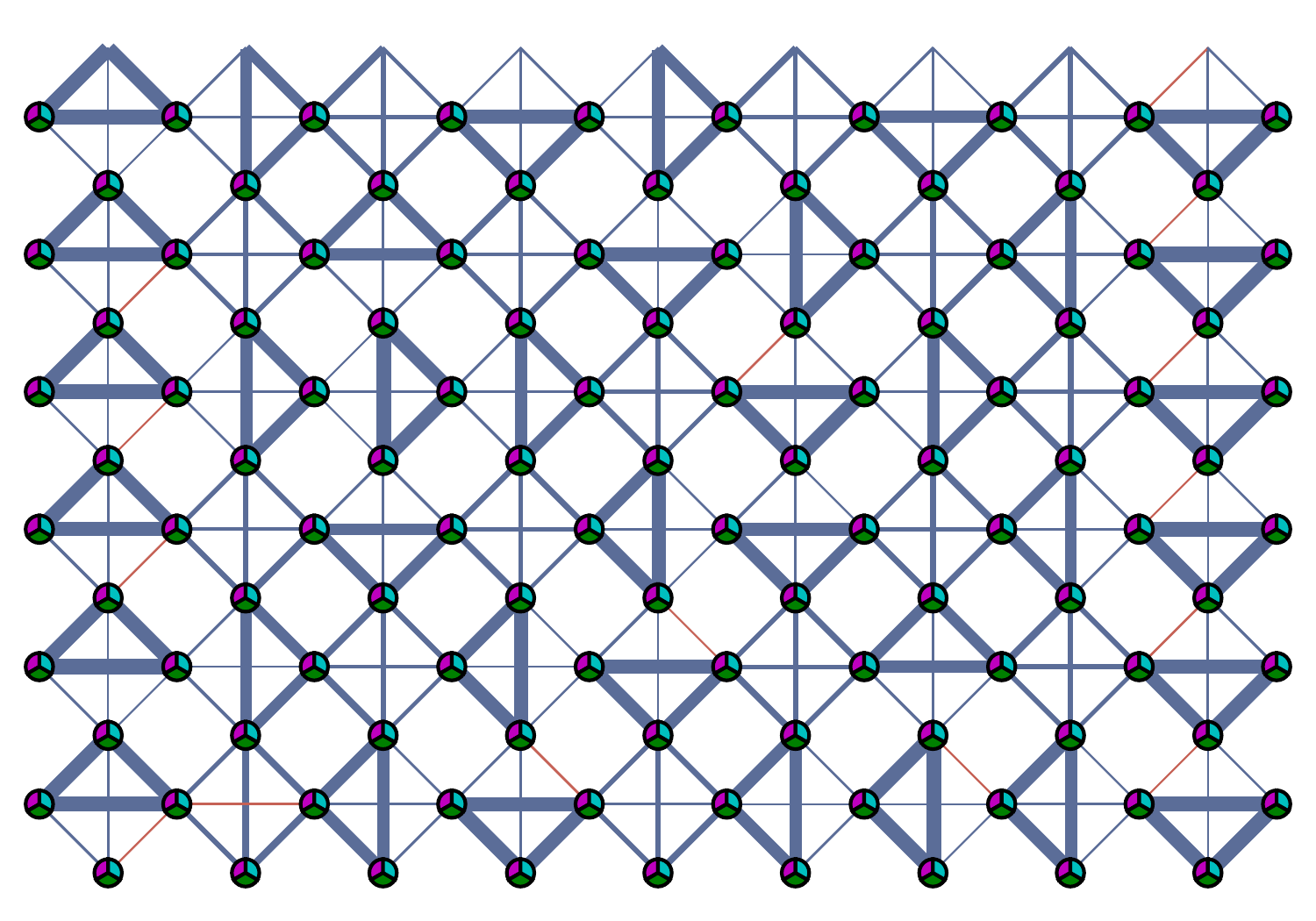}}
  \\
  \subfloat[$\left(10,6\right)$ triangle pattern A\label{fig:local_unitcell_6xx_e}]{
      \includegraphics[width=0.40\textwidth]{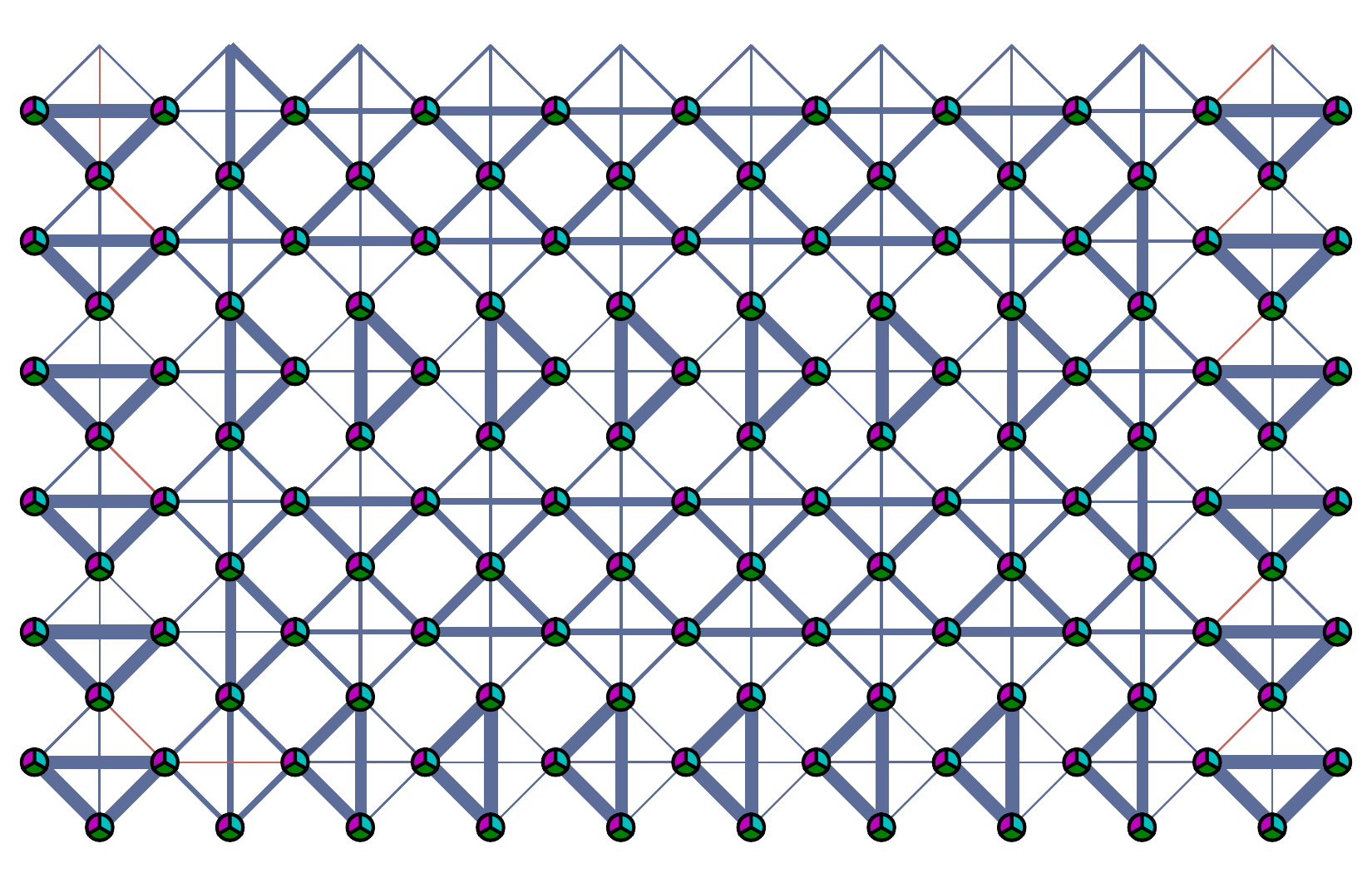}}
  \hspace{0.05\textwidth}
  \subfloat[$\left(10,6\right)$ stripe pattern\label{fig:local_unitcell_6xx_f}]{
      \includegraphics[width=0.40\textwidth]{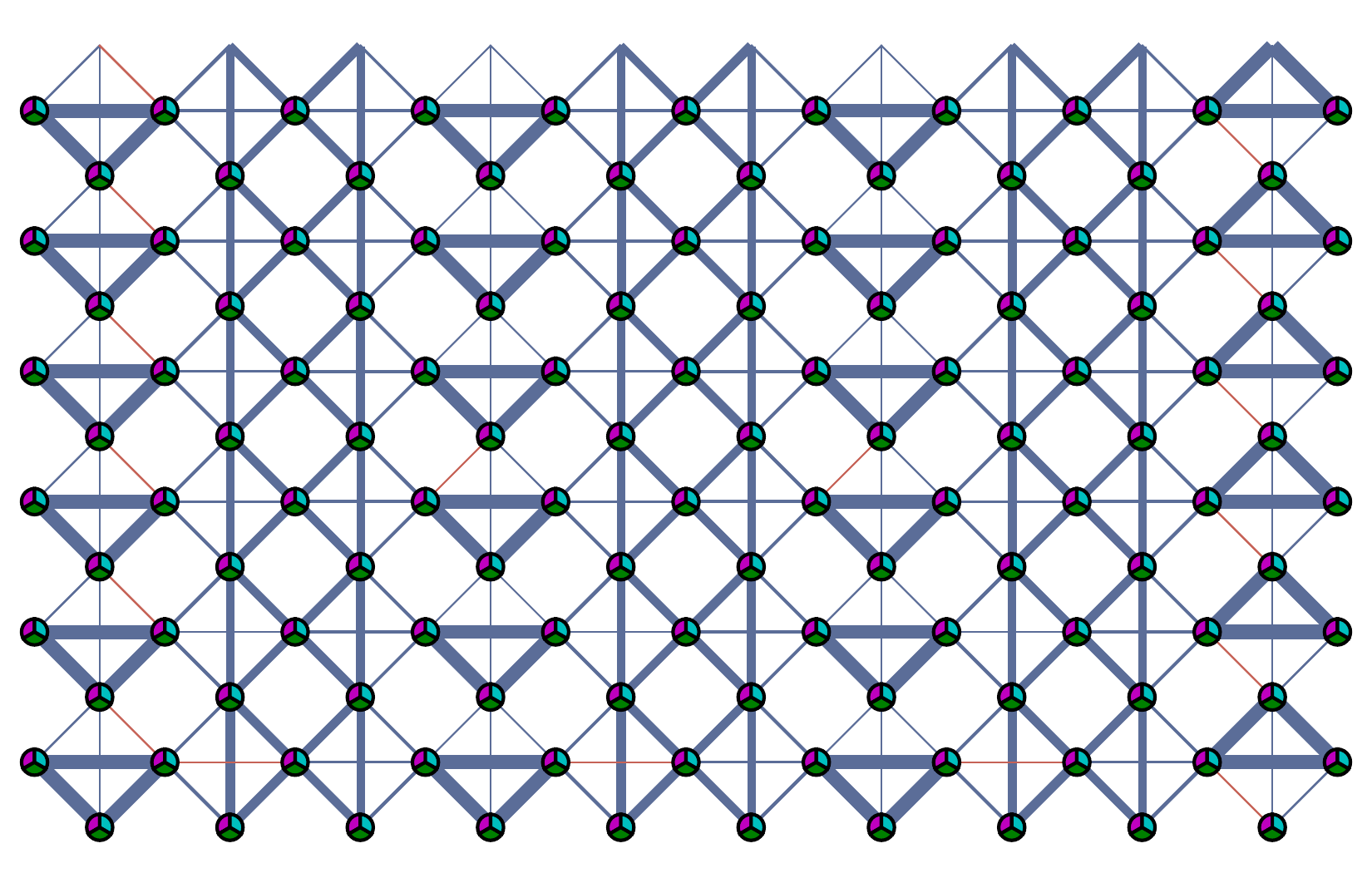}}
      \captionsetup{justification=raggedright,singlelinecheck=false}
      \caption[local_unitcell_6xx]{%
  Ground states and local minimum states obtained by DMRG simulation for the system width $N_y=6$. %
  The energy of the states are  (\ref{fig:local_unitcell_6xx_a}) $E_0 = -342.361$, %
                                (\ref{fig:local_unitcell_6xx_b}) $E_0 = -384.193$, %
                                (\ref{fig:local_unitcell_6xx_c}) $E_0 = -384.550$, %
                                (\ref{fig:local_unitcell_6xx_d}) $E_0 = -384.070$, %
                                (\ref{fig:local_unitcell_6xx_e}) $E_0 = -426.819$, %
                                (\ref{fig:local_unitcell_6xx_f}) $E_0 = -426.903$. %
  The bond stripe states have better energy. %
  }\label{fig:local_unitcell_6xx}
\end{figure}

When the system size in the $y$ direction is 6, we can freely adjust the size in the $x$ direction, ensuring the total number of spins is a multiple of 3.
The results are shown in FIG.~\ref{fig:local_unitcell_6xx}.

\section{Low energy effective theory of local triangle singlet}
\begin{figure}[h!]
    \centering
        \includegraphics[width=0.40\textwidth]{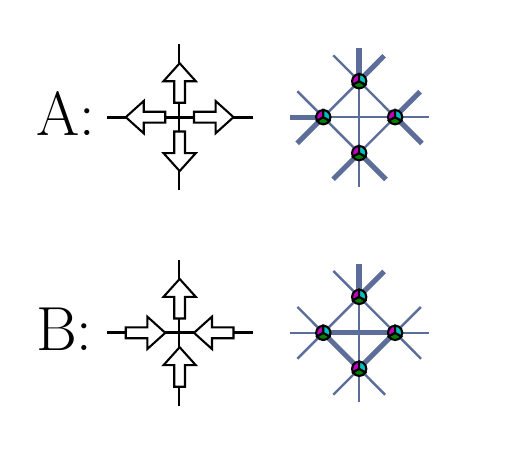}
        \captionsetup{justification=raggedright}
        \caption[arrow_rep_map]{%
        Mapping for the local triangle singlet to the arrow representation: case A represents four spins in the unit cell to form a 3-spin singlet in its surrounding unit cell; %
        case B represents three spins in the unit cell form a $\rm{SU}\left(3\right)$ singlet, and one spin forms a singlet in the adjacent unit cell.
    }\label{fig:arrow_rep_map}
\end{figure}

\begin{figure}[h!]
    \centering
    \includegraphics[width=0.40\textwidth]{figs/local_kiteladder_4x7_tri.pdf}
    \hspace{0.05\textwidth}
    \includegraphics[width=0.40\textwidth]{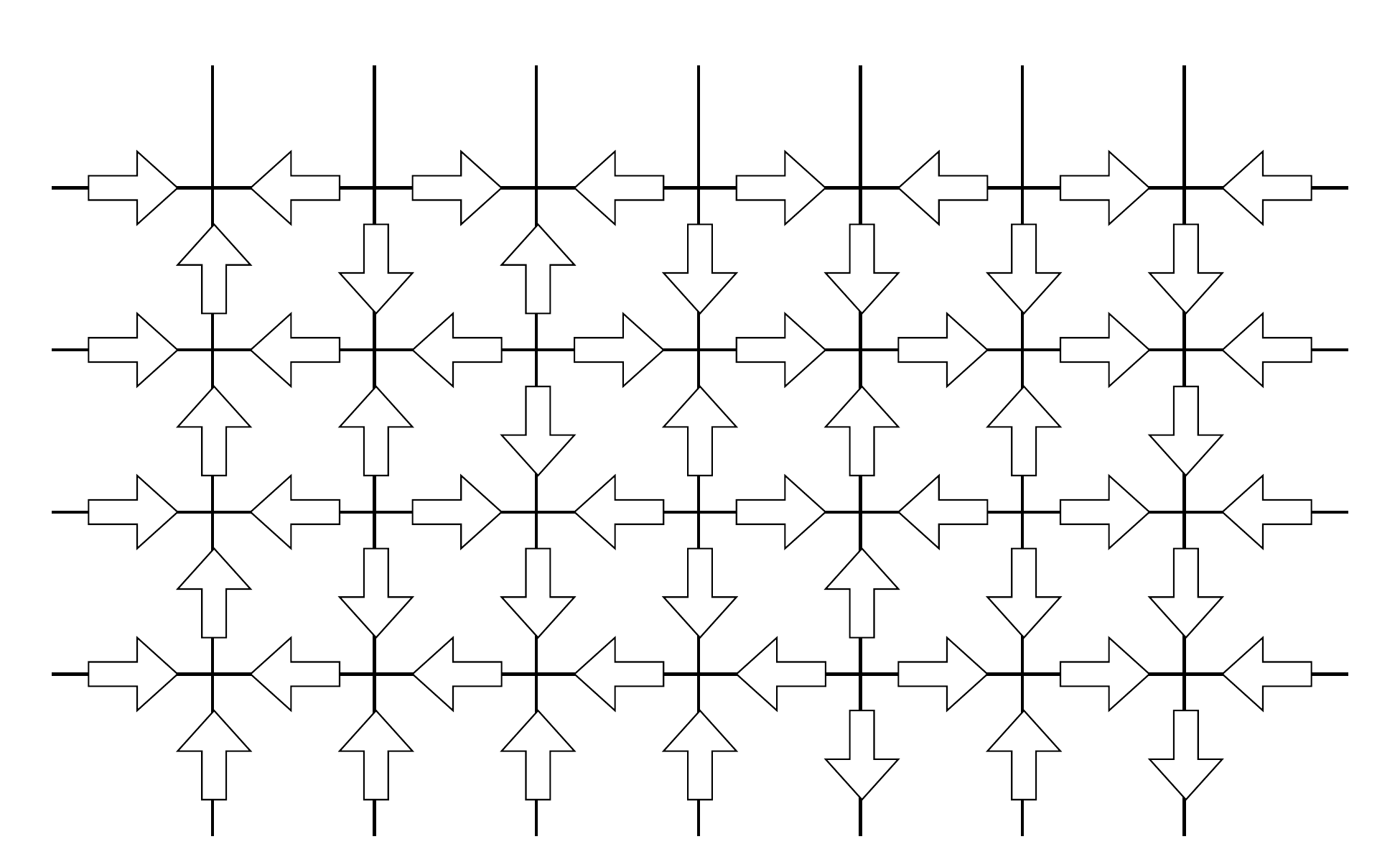}
    \captionsetup{justification=raggedright}
    \caption[4x7_arrow_mapping]{%
    An example of the local triangle singlet for the system size $\left(4,7\right)$ and its arrow representation. %
        }
    \label{fig:arrow_rep_4x7}
\end{figure}

The arrow mapping rule is illustrated in FIG.~\ref{fig:arrow_rep_map}, and a real-space example is provided in FIG.~\ref{fig:arrow_rep_4x7}.

Through the mapping, the local triangle singlet pattern in the low-energy effective theory corresponds to a restricted vertex model space.
This vertex model space contains only sources and no sinks.
The flow of arrows is not constant; instead, it forms loops.
In previous finite-size DMRG calculations, we observed that the boundary configurations exhibit loop structures.

\begin{figure}[h!]
    \centering
    \includegraphics[width=0.45\textwidth]{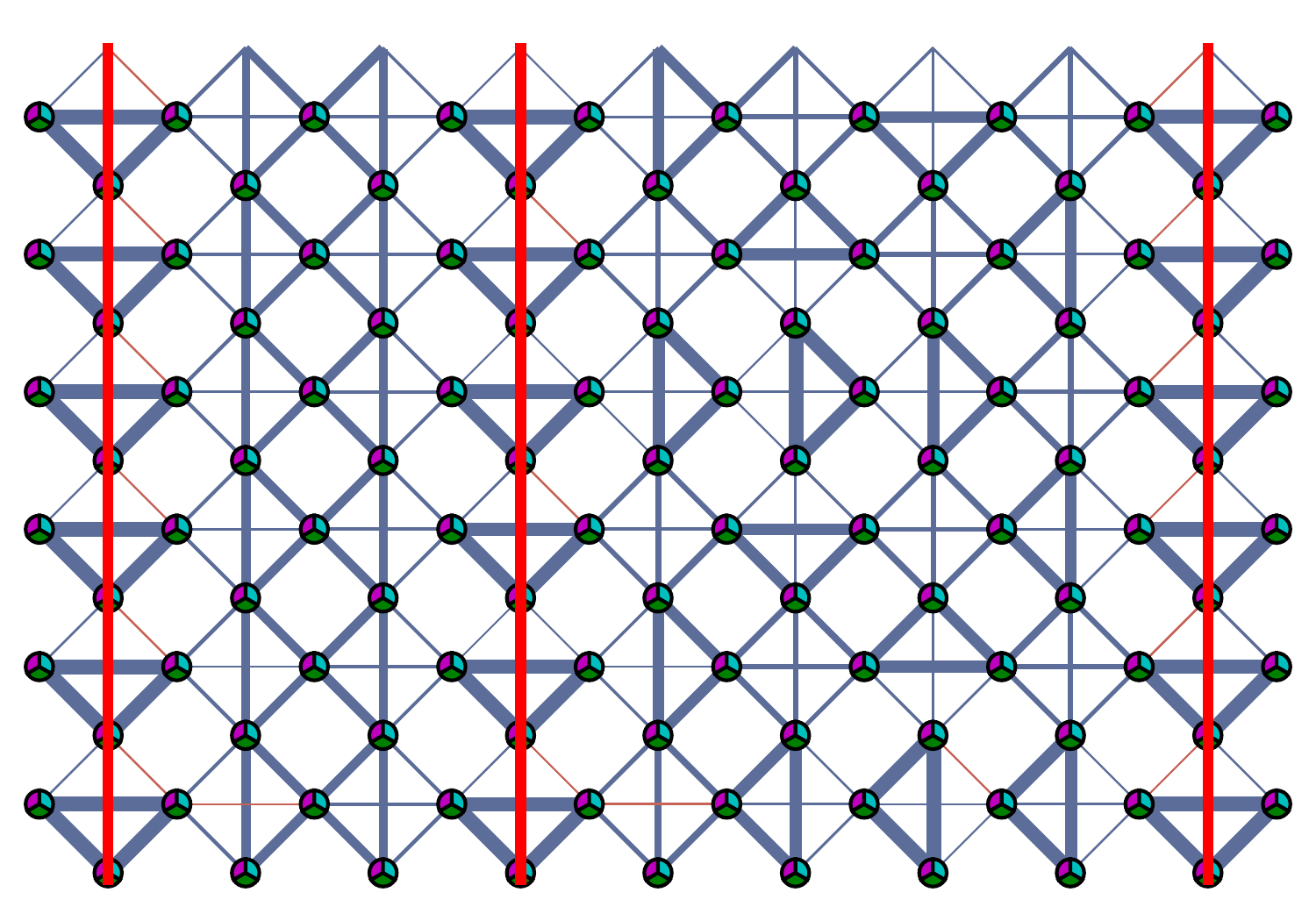}
    \hspace{0.05\textwidth}
    \includegraphics[width=0.45\textwidth]{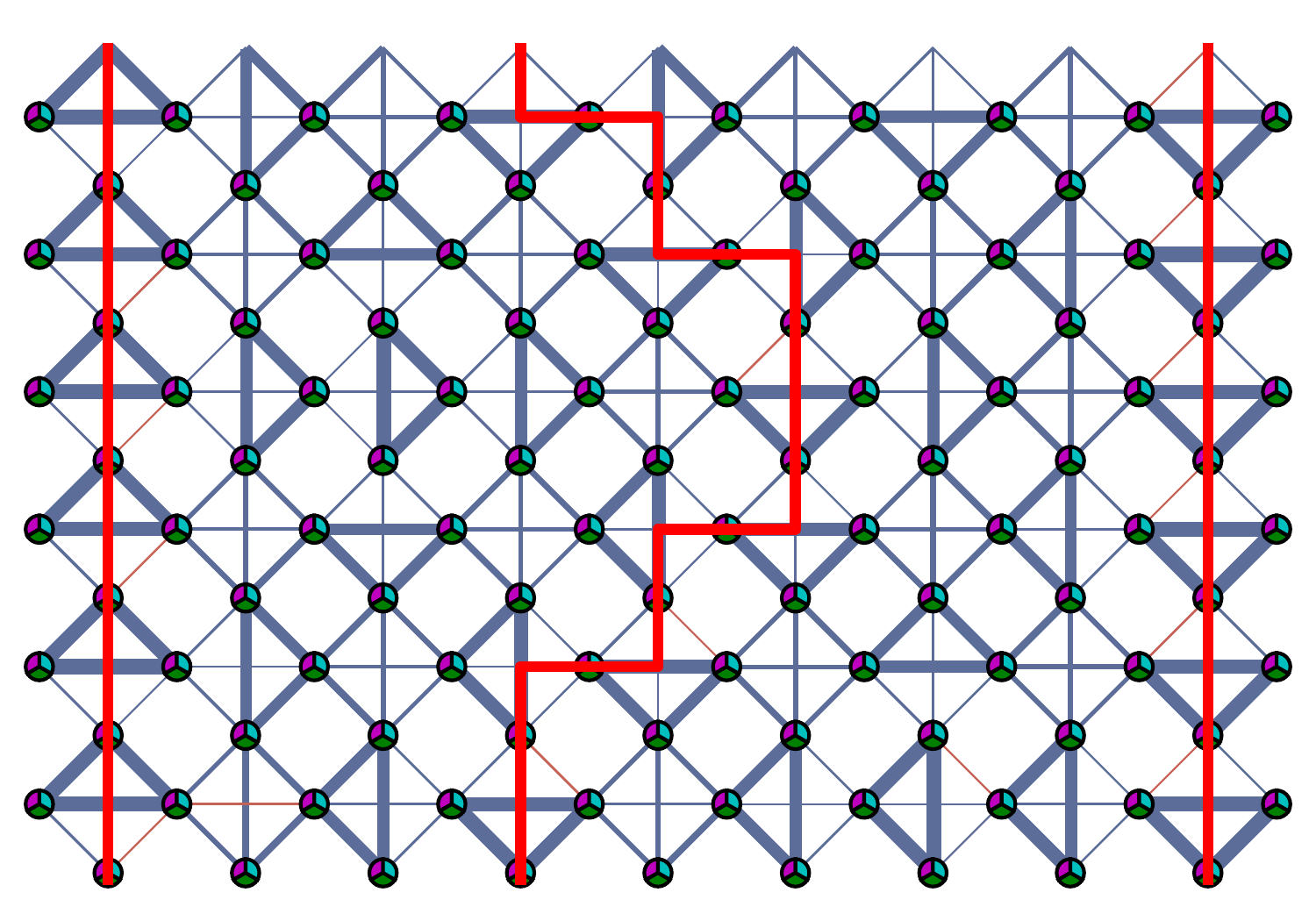}
    \caption[two_loop_9x6]{%
        Two loop structures for the system size $\left(N_x,N_y\right)=\left(9,6\right)$, labeled by red lines. %
        The reverse operation on the arrows also corresponds to another real space configuration. %
    }\label{fig:9x6_loop}
\end{figure}

As shown in FIG.~\ref{fig:9x6_loop}, for the system size $\left(N_x,N_y\right)=\left(9,6\right)$, there are two internal structures.
Both are characterized by an in-body loop structure.

\end{document}